\documentclass[a4paper,11pt]{article} 

\usepackage{amsmath}
\usepackage{latexsym}
\usepackage{amsfonts}
\usepackage{graphicx} 
\usepackage{subfigure}
\usepackage{amssymb}
\usepackage{stmaryrd}
\usepackage{caption}
\usepackage{color}
\usepackage{enumerate}

\newcommand{\dfn}{\triangleq}
\newcommand{\QED}{\Box} 
\newcommand{\rw}{\rightarrow} 
\newcommand{\Prob}{\mathbb{P}}    
\newcommand{\Real}{\mathbb{R}}  
  
\newcommand{\mB}{{\mathcal B}}
\newcommand{\mP}{{\mathcal P}}

\newcommand{\mC}{{\mathcal C}}

\newcommand{\mS}{{\mathcal S}}
\newcommand{\mI}{{\mathcal I}}

\newcommand{\mG}{{\mathcal G}}

\newcommand{\mX}{{\mathcal X}}
\newcommand{\mN}{{\mathcal N}}
\newcommand{\mU}{{\mathcal U}}

\newcommand{\sfC}{{\sf C}}
\newcommand{\sfk}{{\sf k}}
\newcommand{\sfK}{{\sf K}}

\newcommand{\sfp}{{\sf p}}

\newtheorem{Teorema}{\em Theorem}
\newtheorem{Definicion}{\em Definition}
\newtheorem{Lema}{\em Lemma}
\newtheorem{Proposicion}{\em Proposition}

\newtheorem{Assumption}{\em A.}
\newtheorem{Nota}{\em Remark}
\newtheorem{Algoritmo}{\em Algorithm}

\newtheorem{Result}{\em Result}

\title{Uniform convergence over time of a nested particle filtering scheme for recursive parameter estimation in state--space Markov models}

\author{Dan Crisan\thanks{Department of Mathematics, Imperial College London (UK). E-mail: {\tt d.crisan@imperial.ac.uk}.} \and Joaqu\'{\i}n M\'{\i}guez\thanks{School of Mathematical Sciences, Queen Mary University of London (UK). E-mail: {\tt j.miguez@qmul.ac.uk}.}}
 
\begin{document}

\maketitle

\abstract{
We analyse the performance of a recursive Monte Carlo method for the Bayesian estimation of the static parameters of a discrete--time state--space Markov model. The algorithm employs two layers of particle filters to approximate the posterior probability distribution of the model parameters. In particular, the first layer yields an empirical distribution of samples on the parameter space, while the filters in the second layer are auxiliary devices to approximate the (analytically intractable) likelihood of the parameters. This approach relates the this algorithm to the recent sequential Monte Carlo square (SMC$^2$) method, which provides a {\em non-recursive} solution to the same problem. In this paper, we investigate the approximation, via the proposed scheme, of integrals of real bounded functions with respect to the posterior distribution of the system parameters. Under assumptions related to the compactness of the parameter support and the stability and continuity of the sequence of posterior distributions for the state--space model, we prove that the $L_p$ norms of the approximation errors vanish asymptotically (as the number of Monte Carlo samples generated by the algorithm increases) and uniformly over time. We also prove that, under the same assumptions, the proposed scheme can asymptotically identify the parameter values for a class of models. We conclude the paper with a numerical example that illustrates the uniform convergence results by exploring the accuracy and stability of the proposed algorithm operating with long sequences of observations. 
}

\addtolength{\baselineskip}{0.2cm}

\newpage

\section{Introduction}




%

The problem of parameter estimation arises in a multitude of applications of state--space dynamic models and, as a consequence, has received considerable attention from different perspectives \cite{Kitagawa98,Liu01b,Andrieu04,Kantas09,Carvalho10,Kantas15}. We investigate the use of a nested particle filtering scheme, introduced in \cite{Crisan15}, for the recursive Bayesian estimation of the static parameters of discrete-time state-space Markov systems. 

\subsection{Background}

To ease the presentation, let us consider two (possibly vector-valued) random sequences $\{ X_t \}_{t=0,1,...}$ and $\{ Y_t \}_{t=1,2,...}$ representing the (hidden) state of a dynamic system and some related observations, respectively, with $t$ denoting discrete time. The state process is assumed to be Markov and the observation $Y_t$ is independent of any other observations $\{ Y_k; k \ne t \}$, conditional on the state $X_t$. The conditional probability distribution of $X_t$ given $X_{t-1}=x_{t-1}$ and the probability density function (pdf) of $Y_t$ given $X_t=x_t$ are assumed to be known up to a vector of static random parameters, denoted  $\Theta$. These assumptions are very common in the literature and actually hold for many practical systems (see, e.g., \cite{Ristic04,Cappe07}). Given a sequence of observations, $Y_1 = y_1, \ldots, Y_t = y_t, \ldots$, the Bayesian parameter estimation problem consists in tracking the posterior probability distribution of the parameter vector $\Theta$ over time.

When the parameter vector is known, $\Theta=\theta$, it is a common approach to use particle filters  \cite{Gordon93,Kitagawa96,Liu98,Doucet00,Pitt01,Doucet01b,Ristic04,Cappe07,Kunsch13} in order to track (over time $t$) the posterior probability distribution of the state $X_t$ conditional the record of observations, $Y_{1:t}=y_{1:t}$, which is often termed the {\em filtering distribution}. At each time step, a particle filter generates a discrete random approximation of the filtering distribution that consists of samples on the state space. Unfortunately, the design of particle filtering methods that can account for a random vector of parameters in the dynamic model (i.e., a static but unknown $\Theta$) is a hard problem and it has remained an open issue for two decades. While many algorithms have been proposed \cite{LeGland97,Chen00b,Liu01b,Storvik02,Andrieu04,Papavasiliou06,Carvalho10,Poyiadjis11} none of them is widely accepted as a complete solution to this problem. Some of them are seen as {\em ad hoc} \cite{Liu01b}, others depend on the structure of the state--space model to be applicable \cite{Chen00b,Storvik02,Carvalho10} and others yield only point estimates rather than approximations of the sequence of posterior distributions \cite{LeGland97,Andrieu04,Poyiadjis11}.  The recent sequential Monte Carlo square (SMC$^2$) method \cite{Chopin12} overcomes these problems, but the algorithm is {\em not} recursive and hence it becomes computationally prohibitive when the sequence of observations is relatively long. See \cite{Kantas15} for a recent survey of the field.

\subsection{Contributions}

We investigate the convergence and performance of the nested particle filtering scheme in \cite{Crisan15} for the approximation of the posterior distribution of the unknown parameters $\Theta$ given the data $Y_{1:t}=y_{1:t}$. Similar to \cite{Papavasiliou06} and \cite{Chopin12}, the algorithm consists of two nested layers of particle filters: an ``outer'' filter that approximates the probability measure of $\Theta$ given the observations and a bank of ``inner'' filters that yield approximations of the posterior probability distribution of $X_t$ conditional on specific realisations of $\Theta$. The outer filter directly provides an approximation of the marginal posterior distribution of $\Theta$, which is the main object of interest in this paper. The proposed scheme is similar to the 
SMC$^2$ method of \cite{Chopin12}. However, unlike SMC$^2$, it is a purely recursive procedure that readily admits an online implementation. A detailed comparison of the two algorithms is provided in \cite{Crisan15}.

In this paper we look into the approximation, via the proposed scheme, of integrals of real bounded functions with respect to (w.r.t.) the posterior distribution of the system parameters. Under a set of assumptions related to
\begin{itemize}
\item the compactness of the parameter space,
\item the stability of the sequence of posterior probability measures associated to $\Theta$ and $X_t$, and
\item the continuity of the conditional (on $\Theta$) optimal filters in the state--space
\end{itemize}
we prove that the $L_p$ norms of the approximation errors vanish asymptotically, as the number of particles in the filter increases, and uniformly over time. In particular, we obtain an explicit upper bound for the $L_p$ approximation errors that is independent of the time index $t$. This uniform convergence result has some relevant consequences. One of them is that the proposed scheme can eventually identify the parameter values for a broad class of state-space models. In particular, we prove that, when the true posterior probability measure of $\Theta$ converges toward a unit delta measure located at a point $\theta_*$ in the parameter space, the approximation computed via the proposed nested particle filter also converges to the same delta, in terms of a suitable distance, as $t\rw\infty$. 

In order to illustrate the theoretical results, we present computer simulation results, for a stochastic Lorenz 63 model, which show numerically how the nested particle filtering algorithm attains an accurate and stable performance with a fixed number of particles and long sequences of observations.

\subsection{Organisation of the paper}
 
We present a general description of the random state-space Markov models of interest in this paper in Section \ref{sModel}. In Section \ref{sNested} we describe the proposed nested particle filtering scheme. A summary of the theoretical findings in the paper is provided in Section \ref{sResults}, while the full analysis of the algorithm is described in Section \ref{sUniform}. In Section \ref{sExamples} we 
present the results of our computer simulation experiments. Finally, Section \ref{sConclusions} is devoted to the conclusions.

\section{Background} \label{sModel}

\subsection{Notation, assumptions and preliminary results}

We first introduce some common notations to be used through the paper, broadly classified by topics. Below, $\Real$ denotes the real line, while for an integer $d\ge 1$, $\Real^d=\overbrace{\Real \times \ldots \times \Real}^{d \mbox{ {\tiny times}}}$

\begin{itemize}

\item Functions: Let $S \subseteq \Real^d$ be a subset of $\Real^d$.
        \begin{itemize}
        \item The supremum norm of a real function $f:S \rw \Real$ is denoted as $\| f \|_\infty = \sup_{x\in S} | f(x) |$.
        \item $B(S)$ is the set of bounded real functions over $S$, i.e., $f \in B(S)$ if, and only if, $\| f \|_\infty < \infty$.
	\item We use $a \vee b$ and $a \wedge b$ to denote the maximum and the minimum, respectively, between two real numbers $a$ and $b$.
        \end{itemize}
        
\item Measures and integrals:
        \begin{itemize}
        \item $\mB(S)$ is the $\sigma$-algebra of Borel subsets of $S$.
        \item $\mP(S)$ is the set of probability measures over the measurable space $(S,\mB(S))$.
        \item $(f,\mu) \dfn \int f(x) \mu(dx)$ is the integral of a real function $f:S \rw \Real$ w.r.t. a measure $\mu \in \mP(S)$.
        \item Given a probability measure $\mu \in \mP(S)$, a Borel set $A \in \mB(S)$ and the indicator function 
        $$
        I_A(x) = \left\{
                \begin{array}{ll}
                1, &\mbox{if } x \in A\\
                0, &\mbox{otherwise}
                \end{array}
        \right.,
        $$
        $\mu(A) = (I_A,\mu) = \int I_A(x) \mu(dx)$ 
        is the probability of $A$.

        \end{itemize}

        
\item Sequences, vectors and random variables (r.v.'s):
        \begin{itemize}
        \item We use a subscript notation for sequences, namely $x_{t_1:t_2} \dfn \{ x_{t_1}, \ldots, x_{t_2} \}$.
        \item For an element $x=(x_1,\ldots,x_d) \in \Real^d$, its Euclidean norm is denoted as $\| x \| = \sqrt{ x_1^2+\ldots+x_d^2 }$.
        \item The $L_p$ norm of a real r.v. $Z$, with $p \ge 1$, is written as $\| Z \|_p \dfn E[ |Z|^p ]^{1/p}$, where $E[\cdot]$ denotes expectation w.r.t. the probability distribution of $Z$.
        \end{itemize}
\end{itemize}

\begin{Nota}
Let $\alpha, \beta, \bar \alpha, \bar \beta \in \mP(S)$ be probability measures and let $f,h \in B(S)$ be two real bounded functions on $S$ such that $(h,\bar \alpha)>0$ and $(h,\bar \beta)>0$. If the identities 
$$
(f,\alpha) = \frac{
	(fh,\bar \alpha)
}{
	(h,\bar \alpha)
} \quad \mbox{and} \quad (f,\beta) = \frac{
	(fh,\bar \beta)
}{
	(h,\bar \beta)
}
$$
hold, then it is straightforward to show (see, e.g., \cite{Crisan01}) that
\begin{equation}
| (f,\alpha)-(f,\beta) | \le \frac{
	1
}{
	(h,\bar \alpha)
} \left|
	(fh,\bar \alpha) - (fh,\bar \beta)
\right| + \frac{
	\| f \|_\infty
}{
	(h,\bar \alpha)
} \left|
	(h,\bar \alpha) - (h,\bar \beta)
\right|. 
\label{eqPreliminaries}
\end{equation}
\end{Nota}

\subsection{State-space Markov models in discrete time} \label{ssModel}

Consider two random sequences, $\{ X_t \in \mX \}_{t \ge 0}$ and $\{ Y_t \in \Real^{d_y} \}_{t \ge 1}$, and a random variable $\Theta \in D_\theta$, where $\mX \subseteq \Real^{d_x}$, $D_\theta \subset \Real^{d_\theta}$ and the positive integers $d_x$, $d_y$ and $d_\theta$ determine the dimension of the state space, the observation space and the parameter space, respectively. We further assume that $D_\theta$ is compact. Let $\Prob_t$ be the joint probability measure for the triple $\left( \{X_n\}_{n \le t}, \{Y_n\}_{1 \le n \le t}, \Theta \right)$, that we assume to be absolutely continuous w.r.t. the Lebesgue measure. 

The  sequence $\{ X_t \}_{t\ge 0}$ is the state (or signal) process, a possibly inhomogeneous Markov chain governed by an initial probability measure $\tau_{0} \in \mP(\mX)$ and a sequence of transition kernels $\tau_{t,\theta} : \mB(\mX) \times \mX \rw [0,1]$ indexed by a realisation of the r.v. $\Theta=\theta$. To be specific, we define
\begin{eqnarray}
\tau_{0}(A) &\dfn& \Prob_0\left\{ X_0 \in A \right\}, \label{eqPrior} \\ 
\tau_{t,\theta}(A|x_{t-1}) &\dfn& \Prob_t\left\{ X_t \in A | X_{t-1}=x_{t-1}, \Theta=\theta \right\}, \quad t \ge 1, \label{eqKernel}
\end{eqnarray}
where $A \in \mX$ is a Borel set. The sequence $\{ Y_t \}_{t \ge 1}$ is termed the observation process. Each r.v. $Y_t$ is assumed to be conditionally independent of other observations given $X_t$ and $\Theta$, namely
\begin{equation}
\Prob_t\left\{ Y_t \in A | X_{0:t} = x_{0:t}, \Theta = \theta, \{ Y_k = y_k \}_{k \ne t} \right\} = \Prob_t\left\{ Y_t \in A | X_t = x_t, \Theta=\theta  \right\} \nonumber
\end{equation}
for any $A \in \mB(\Real^{d_y})$. Additionally, we assume that, for every $x \in \mX$ and $\theta \in D_\theta$, the r.v. $Y_t|X_t=x,\Theta=\theta$ has an associated probability density function (pdf). In particular, for some possibly unknown normalisation constant $c$, there are functions $g_{t,\theta}(y|x)$ such that
\begin{equation}
\Prob_t\left\{ Y_t \in A | X_t = x_t, \Theta=\theta  \right\}  = c \int I_A(y) g_{t,\theta}(y|x_t)dy.
\nonumber
\end{equation}
We assume that $c$ is independent of $y$, $x$ and $\theta$.  

If $\Theta=\theta$ (the parameter is given), then the stochastic filtering problem consists in the computation of the posterior probability measure of the state $X_t$ given the parameter and a sequence of observations up to time $t$. Specifically, for a given observation record $\{ y_t \}_{t \ge 1}$, we seek the measures
\begin{equation}
\phi_{t,\theta}(A) \dfn \Prob_t\left\{ X_t \in A | Y_{1:t}=y_{1:t}, \Theta=\theta \right\}, \quad t=0, 1, 2, ...
\nonumber
\end{equation}
where $A \in \mX$. For many practical applications, the interest actually lies in the computation of integrals of the form $(f,\phi_{t,\theta})$ for some integrable function $f:\mX\rw\Real$. Note that, for $t=0$, we recover the prior signal measure, i.e., $\phi_{0,\theta}=\tau_{0}$ independently of $\theta$. 

We also introduce the predictive measure 
$$
\xi_{t,\theta}(A) \dfn \Prob_t\left\{ X_t \in A | Y_{1:t-1}=y_{1:t-1}, \Theta=\theta \right\}, \quad t=0, 1, 2, ...,
$$ 
which is closely related to the filter $\phi_{t,\theta}$ and we often write as $\xi_{t,\theta} = \tau_{t,\theta}\phi_{t-1,\theta}$, meaning that, for any integrable function $f:\mX\rw\Real$, we obtain
\begin{equation}
(f,\xi_{t,\theta}) = \int \int f(x)\tau_{t,\theta}(dx|x')\phi_{t-1,\theta}(dx') 
=  \left(
	(f,\tau_{t,\theta}), \phi_{t-1,\theta}
\right).
\label{eqPredicting}
\end{equation}
Let us note that $\int f(x)\tau_{t,\theta}(dx|x')$ is itself a map $\mX\rw\Real$. Integrals w.r.t. the filter measure $\phi_{t,\theta}$ can be rewritten by way of $\xi_{t,\theta}$ as 
\begin{equation}
(f,\phi_{t,\theta}) = \frac{
	(fg_{t,\theta}^{y_t},\xi_{t,\theta})
}{
	(g_{t,\theta}^{y_t},\xi_{t,\theta})
},
\label{eqAlternative}
\end{equation}
where $g_{t,\theta}^{y_t}(x) \dfn g_{t,\theta}(y_t|x)$ is the likelihood of $x$. Eqs. \eqref{eqPredicting} and \eqref{eqAlternative} are used extensively through the rest paper.

In the sequel, we assume the parameter $\Theta$ is unknown and focus on the problem of approximating the sequence of probability measures
$$
\mu_t(A) \dfn \Prob_t\left\{ \Theta \in A | Y_{1:t} = y_{1:t} \right\}, \quad t = 0, 1, 2, ..., \mbox{ where } A \in \mB(D_\theta)
$$
that result from the state--space Markov model and the sequence of observations $\{y_{1:t}\}_{t\ge 1}$.

\section{Nested particle filtering algorithm} \label{sNested}

\subsection{Recursive decomposition of $\mu_t$}

Assume that the observations $Y_{1:t-1}=y_{1:t-1}$ are fixed and let
\begin{equation}
\upsilon_{t,\theta}(A) = \Prob_t\left\{ 
	Y_t \in A | Y_{1:t-1}=y_{1:t-1}, \Theta = \theta 
\right\}, \quad A \in \mB(\Real^{d_y}),
\end{equation}
be the probability measure associated to the (random) observation $Y_t$ given $Y_{1:t-1}=y_{1:t-1}$ and the parameter vector $\Theta=\theta$. Let us assume that $\upsilon_{t,\theta}$ has a density $u_{t,\theta}:\Real^{d_y} \rw [0,+\infty)$ w.r.t. the Lebesgue measure, i.e., for any $A \in \mB(\Real^{d_y})$,
\begin{equation}
\upsilon_{t,\theta}(A) = \int I_A(y) u_{t,\theta}(y) dy.
\nonumber
\end{equation}
The posterior probability measure of the parameter, $\mu_t$, can be related to the predictive measure $\xi_{t,\theta}$ by way of the pdf $u_{t,\theta}(y)$. To be precise, for given $Y_t=y_t$ and $\Theta=\theta$, the density $u_{t,\theta}(y_t)$ can be written as the integral
$$
u_{t,\theta}(y_t) = (g_{t,\theta}^{y_t}, \xi_{t,\theta}),
$$ 
which yields the marginal likelihood of the parameter value $\theta$, denoted in the sequel as
$$
u_t(\theta) \dfn u_{t,\theta}(y_t) = (g_{t,\theta}^{y_t}, \xi_{t,\theta}).
$$ 
Then, it is a straightforward application of Bayes' theorem to show that the sequence of measures $\mu_t$ obeys the recursion
\begin{equation}
(h,\mu_t) = \frac{
	(hu_t,\mu_{t-1})
}{
	(u_t,\mu_{t-1})
}, \quad \mbox{for $t=1, 2, ...$}
\label{eqRecursion_mu_t}
\end{equation}
for any integrable function $h:D_\theta\rw\Real$.

Equation \eqref{eqRecursion_mu_t} suggests the implementation of a sequential Monte Carlo (SMC) approximation of $\mu_t$. In particular, at time $t$ one could 
\begin{itemize}
\item draw $N$ i.i.d. samples $\{ \bar \theta_t^{(i)} \}_{1\le i \le N}$ from the posterior measure at time $t-1$, $\mu_{t-1}$, 
\item and then compute normalised importance weights proportional to the marginal likelihoods $u_t(\bar \theta_t^{(i)})$.
\end{itemize}
However, neither sampling from $\mu_{t-1}$ nor the computation of the likelihood $u_t(\theta)$ can be carried out exactly, hence some approximations are needed. This is explored in Subsections \ref{ssJitteringKernel} and \ref{ssLikelihoodTheta}, respectively.

\subsection{Sampling in the parameter space} \label{ssJitteringKernel}

Assume that a particle approximation $\mu_{t-1}^N = \frac{1}{N}\sum_{i=1}^N \delta_{\theta_{t-1}^{(i)}}$ of $\mu_{t-1}$ is available. A natural way to generate a new sample of size $N$ distributed approximately as $\mu_{t-1}$ is to {\em jitter} the particles $\theta_{t-1}^{(1)}, ..., \theta_{t-1}^{(N)}$. 
\begin{Nota} \label{rmSMC}
This random jittering, or rejuvenation, of the particles in the parameter space is necessary in order to avoid the degeneracy of the SMC method \cite{Liu01b}, but the error introduced by this step should be controlled. In the SMC$^2$ framework of \cite{Chopin12}, this is done by applying a particle Markov chain Monte Carlo (pMCMC) kernel to the particle set $\{ \theta_{t-1}^{(i)} \}_{i=1}^N$ that leaves its underlying distribution invariant. However, this procedure implies the processing of the complete sequence of observations up to time $t$, ${\bf y}_{1:t}$, and, therefore, prevents a recursive implementation.
\end{Nota}

To circumvent the drawback described in Remark \ref{rmSMC}, we propose to use Markov kernels of the form 
\begin{equation}
\kappa_{N,\sfp}^{\theta_{t-1}^{(i)}}(d\theta) = (1-\epsilon_{N,\sfp})\delta_{\theta_{t-1}^{(i)}}(d\theta) + \epsilon_{N,\sfp} \bar \kappa^{\theta_{t-1}^{(i)}}(\theta)d\theta, \quad i=1, 2, \ldots, N,
\label{eqKappa}
\end{equation}
where $\epsilon_{N,\sfp} \in \left( 0,\frac{1}{N^\frac{\sfp}{2}} \right]$, $\sfp \ge 1$, and $\bar \kappa^{\theta'}(\theta)$ is a pdf w.r.t. the Lebesgue measure, independent of $N$, centred at $\theta'$ and with support in $D_\theta$, i.e., $\int \theta \bar \kappa^{\theta'}(\theta)d\theta = \theta'$ and $\int I_{D_\theta}(\theta) \bar \kappa^{\theta'}(\theta)d\theta = 1$. It is relatively straightforward to show that kernels in the class described by \eqref{eqKappa} satisfy the inequalities stated below.

\begin{Proposicion} \label{propAssumptionKappa1}
If $\kappa_{N,\sfp}$ is selected as in Eq. \eqref{eqKappa}, then
\begin{equation}
\sup_{\theta' \in D_\theta} \int \left|
	h(\theta) - h(\theta')
\right| \kappa_{N,\sfp}^{\theta'}(d\theta) \le \frac{
	2 \| h \|_\infty
}{
	\sqrt{N}
}
\label{eqAssKappa1}
\end{equation}
for any $h \in B(D_\theta)$, and
\begin{equation}
\sup_{\theta' \in D_\theta} \int \left\|
	\theta - \theta'
\right\|^\sfp \kappa_{N,\sfp}^{\theta'}(d\theta) \le \frac{
	c_\kappa^\sfp
}{
	N^{\frac{p}{2}}
},
\label{eqAssKappa2}
\end{equation}
where $c_\kappa < \infty$ is a constant independent of $N$.
\end{Proposicion}

\noindent \textit{\textbf{Proof:}} It is straightforward. Simply note that $|h(\theta) - h(\theta')|\le 2\|h\|_\infty$ to arrive at \eqref{eqAssKappa1}. Inequality \eqref{eqAssKappa2} is readily obtained, with $c_\kappa = \sup_{\theta,\theta'\in D_\theta} \|\theta-\theta'\|<\infty$, if we recall that $D_\theta$ is defined to be compact. $\QED$ 

See \cite[Section 5.1]{Crisan15} for a more detailed discussion of the choice of the jittering kernel, including some variations on the family of equation \eqref{eqKappa}. In the sequel, we assume that $\kappa_{N,\sfp}^{\theta_{t-1}^{(i)}}(d\theta)$ is selected according to \eqref{eqKappa}, so that Proposition \ref{propAssumptionKappa1} holds.

\subsection{Approximation of the parameter likelihood function $u_t(\theta)$} \label{ssLikelihoodTheta}

The second ingredient that we need in order to construct a SMC algorithm that approximates the measures $\mu_t$ is a method to compute the likelihood $u_t(\theta)$. For fixed $\Theta=\bar \theta_t^{(i)}$, the value $u_t(\bar\theta_t^{(i)})$ can be estimated using a standard particle filter (or {\em bootstrap} filter \cite{Gordon93}, see also \cite{Doucet01}). This classical algorithm can be  written down (in a convenient form) using the following notation for two random transformations of discrete sample sets on the state space $\mX$.

\begin{Definicion}
Let $\{ x^{(j)} \}_{1 \le j \le M}$ be a set of $M$ points on $\mX$. The random set 
$$\{ \bar x^{(j)} \}_{1 \le j \le M} = \Upsilon_{n,\theta}\left( \{ x^{(j)} \}_{1 \le j \le M} \right)$$ 
is obtained by sampling each $\bar x^{(j)}$ from the corresponding transition kernel $\tau_{n,\theta}(dx|x^{(j)})$, for $j = 1, ..., M$.
\end{Definicion}
\begin{Definicion}
Let $\{ \bar x^{(j)} \}_{1 \le j \le M}$ be a set of $M$ points in $\mX$. The set 
$$\{ x^{(j)} \}_{1 \le j \le M} = \Upsilon_{n,\theta}^{y_n}\left( \{ \bar x^{(j)} \}_{1 \le j \le M} \right)$$ 
is obtained by 
	\begin{itemize}
	\item computing normalised weights proportional to the likelihoods, 
	$$
	v_n^{(j)} = \frac{
		g_{n,\theta}^{y_n}(\bar x_n^{(j)})
	}{
		\sum_{k=1}^M g_{n,\theta}^{y_n}(\bar x_n^{(k)})
	}, \quad j=1, ..., M.
	$$
	\item and then resampling with replacement the set $\{ \bar x^{(j)} \}_{1 \le j \le M}$ according to the weights $\{ v_n^{(j)} \}_{1\le j\le M}$, i.e., assigning $x^{(j)} = \bar x^{(k)}$ with probability $v^{(k)}$, for $j=1, ..., M$ and $k \in \{ 1, ..., M \}$.
	\end{itemize}
\end{Definicion}

The standard particle filter, with $M$ particles per time step and conditional on $\Theta=\theta_t^{(i)}$, can be outlined as follows.

\begin{Algoritmo} \label{alConditionalBF}
Bootstrap filter conditional on $\Theta = \theta_t^{(i)}$.
\begin{enumerate}
\item {\sf Initialisation.} Draw $M$ i.i.d. samples $x_0^{(i,j)}$, $j=1, ..., M$, from the prior distribution $\tau_0$.
\item {\sf Recursive step.} Let $\{ x_{n-1}^{(i,j)} \}_{1 \le j \le M}$ be the set of available samples at time $n-1$, with $n \le t$. The particle set is updated at time $n$ in two steps:
	\begin{enumerate}
	\item Compute $\{ \bar x_n^{(i,j)} \}_{1 \le j \le M} = \Upsilon_{n,\theta_t^{(i)}}\left( \{ x_{n-1}^{(i,j)} \}_{1 \le j \le M} \right)$.
	\item Compute $\{ x_n^{(i,j)} \}_{1 \le j \le M} = \Upsilon_{n,\theta_t^{(i)}}^{y_n}\left( \{ \bar x_n^{(i,j)} \}_{1 \le j \le M} \right)$.
	\end{enumerate}
\end{enumerate}
\end{Algoritmo}

For $n=t$, we obtain random discrete approximations of the posterior probability measures $\xi_{t,\bar \theta_t^{(i)}}(dx_t)$ and $\phi_{t,\bar\theta_t^{(i)}}(dx_t)$ of the form 
\begin{equation}
\xi_{t,\bar \theta_t^{(i)}}^M(dx_t) = \frac{1}{M} \sum_{j=1}^M \delta_{\bar x_t^{(i,j)}}(dx_t) \quad \mbox{and} \quad
\phi_{t,\bar\theta_t^{(i)}}^M(dx_t) = \frac{1}{M} \sum_{j=1}^M \delta_{x_t^{(i,j)}}(dx_t),
\label{eqXiPhi}
\end{equation} 
respectively. Hence, the parameter likelihood $u_t(\bar\theta_t^{(i)}) = ( g_{t,\bar\theta_t^{(i)}}^{y_t},\xi_{t,\bar \theta_t^{(i)}})$, which in general does not have a closed form solution, admits the Monte Carlo approximation 
\begin{equation}
u_t^M(\bar\theta_t^{(i)}) = (g_{t,\bar\theta_t^{(i)}}^{y_t}, \xi_{t,\bar\theta_t^{(i)}}^M) 
= \frac{1}{M} \sum_{j=1}^M g_{t,\bar \theta_t^{(i)}}^{y_t}(\bar x_t^{(i,j)}).
\label{eqApproxLikelihood1}
\end{equation}

\subsection{Nested particle filtering algorithm} \label{ssAlgorithm}

We are now ready to describe the nested particle filtering algorithm which is the main object of analysis in this paper. Essentially, it is a recursive Monte Carlo filter on the parameter space $D_\theta$ that uses conditional bootstrap filters on $\mX$ to approximate the parameter likelihoods. The algorithm is described below.

\begin{Algoritmo} \label{alRA}
Recursive algorithm for the particle approximation of $\mu_t$, $t=0, 1, 2, ...$
\begin{enumerate}
\item {\sf Initialisation.} Draw $N$ i.i.d. samples $\{ \theta_0^{(i)} \}_{1 \le i\le N}$ from the prior distribution $\mu_0(d\theta)$ and $NM$ i.i.d. samples $\{ x_0^{(i,j)} \}_{1 \le i \le N; 1 \le j \le M}$ from the prior distribution $\tau_0$. 

\item {\sf Recursive step.} For $t \ge 1$, assume the particle set $\left\{ \theta_{t-1}^{(i)}, \{ x_{t-1}^{(i,j)} \}_{1 \le j \le M} \right\}_{1 \le i \le N}$
is available and update it taking the following steps.

	\begin{itemize}
	\item[(a)] For each $i=1, ..., N$
		\begin{itemize}
		\item draw $\bar \theta_t^{(i)}$ from  $\kappa_{N,\sfp}^{\theta_{t-1}^{(i)}}(d\theta)$,
		\item update $\{ \bar x_t^{(i,j)} \}_{1 \le j \le M} = \Upsilon_{t,\bar \theta_t^{(i)}}\left(
			\{ x_{t-1}^{(i,j)} \}_{1 \le j \le M}
		\right)$ and construct $\xi_{t,\bar \theta_t^{(i)}}^M = \frac{1}{M}\sum_{j=1}^M \delta_{\bar x_t^{(i,j)}}$,
		\item compute the approximate likelihood $u_t^M(\bar \theta_t^{(i)}) = ( g_{t,\bar \theta_t^{(i)}}^{y_t}, \xi_{t,\bar\theta_t^{(i)}}^M )$, and
		\item update the particle set $\{ \tilde x_t^{(i,j)} \}_{1 \le j \le M} = \Upsilon_{t,\bar \theta_t^{(i)}}^{y_t}\left(
			\{ \bar x_{t}^{(i,j)} \}_{1 \le j \le M}
		\right)$.
		\end{itemize}
	\item[(b)] Compute normalised weights $w_t^{(i)} \propto u_t^M(\bar \theta_t^{(i)})$, $i=1, ..., N$.
	\item[(c)] Resample: for each $i=1, ..., N$, set $\left\{ \theta_t^{(i)},  x_t^{(i,j)} \right\}_{ 1 \le j \le M} = \left\{ \bar \theta_t^{(l)}, \tilde x_t^{(l,j)} \right\}_{1 \le j \le M}$ with probability $w_t^{(l)}$, where $l \in \{ 1, ..., N \}$.
	\end{itemize}
\end{enumerate}
\end{Algoritmo}

Step 2(a) in Algorithm \ref{alRA} involves jittering the samples in the parameter space and then taking a single recursive step of a bank of $N$ standard particle filters. In particular, for each $\bar \theta_t^{(i)}$, $1 \le i \le N$, we have to propagate and resample the particles $\{ x_{t-1}^{(i,j)} \}_{1 \le j \le M}$ so as to obtain a new set $\{ \tilde x_t^{(i,j)} \}_{1 \le j \le M}$. 

\begin{Nota}
The cost of the recursive step in Algorithm \ref{alRA} is independent of $t$. We only have to carry out regular `prediction' and `update' operations in a bank of standard particle filters. Hence, Algorithm \ref{alRA} is sequential, purely recursive and can be implemented online. This is in contrast with the non-recursive (but otherwise similar) SMC$^2$ method of \cite{Chopin12}. A detailed comparison of both techniques is presented in \cite{Crisan15}. 
\end{Nota} 

\begin{Nota}
Algorithm \ref{alRA} yields several Monte Carlo approximations. After the jittering step, we obtain the measure
\begin{equation}
\bar \mu_{t-1}^{N,M} = \frac{1}{N} \sum_{i=1}^N \delta_{\bar \theta_t^{(i)}}
\nonumber
\end{equation}
which is an approximation of $\mu_{t-1}$ computed at time $t$. After the weights are computed at step 2(b), we have the neasure
\begin{equation}
\tilde \mu_t^{N,M} = \sum_{i=1}^N w_t^{(i)} \delta_{\bar \theta_t^{(i)}},
\nonumber
\end{equation}
which approximates the posterior $\mu_t$. After the resampling step 2(c) we have the (unweighted) approximation
\begin{equation}
\mu_t^{N,M} = \frac{1}{N} \sum_{i=1}^N \delta_{\theta_t^{(i)}}
\nonumber
\end{equation}
of $\mu_t$. Conditional predictive and filter measures on the state space are also computed by the inner filters, namely
$$
\xi_{t,\bar \theta_t^{(i)}}^M = \frac{1}{M}\sum_{j=1}^M \delta_{\bar x_t^{(i,j)}}
\quad \mbox{and} \quad
\phi_{t,\theta_t^{(i)}}^M = \frac{1}{M}\sum_{j=1}^M \delta_{x_t^{(i,j)}}.
$$ 
\end{Nota}

\section{Summary of theoretical results} \label{sResults}

In the rest of this paper we look into the particle approximations of the sequence produced by Algorithm \ref{alRA}. For notational simplicity, we assume that the numbers of particles in the inner and outer filters coincide, i.e., $N=M$. Thus, the approximation of the predictive measure $\xi_{t,\bar \theta_t^{(i)}}$ and the filter measure $\phi_{t,\theta_t^{(i)}}$ become $\xi_{t,\bar \theta_t^{(i)}}^N$ and $\phi^N_{t,\theta_t^{(i)}} = \frac{1}{N} \sum_{j=1}^N \delta_{x_t^{(i,j)}}$, respectively. For conciseness, we will also write 
$$
\bar \mu_t^N = \bar \mu_t^{N,N}, \quad \tilde \mu_t^N = \tilde \mu_t^{N,N} \quad \mbox{and} \quad \mu_t^N = \mu_t^{N,N}.
$$ 
The complexity of Algorithm \ref{alRA} with $N=M$ and a sequence of observations of length $T$, $Y_{1:T}=y_{1:T}$, becomes $\mathcal{O}(N^2T)$ \cite{Crisan15}.

While in \cite{Crisan15} we address the consistency of Algorithm \ref{alRA} (as $N,M\rw\infty$) for a finite-length sequence of observations, here we tackle the problem of proving that the proposed nested particle filter actually converges {\em uniformly over time} when the state space model satisfies a set of sufficient conditions. In particular, for the analysis in this paper we assume that
\begin{enumerate}[(i)]
\item the sequence of probability measures $\{\mu_t\}_{t\ge 0}$ is stable w.r.t. its initial value,
\item the Markov kernels $\tau_{t,\theta}(dx_t|x_{t-1})$ are mixing (uniformly, for all $\theta \in D_\theta$) and the likelihood functions $g_{t,\theta}^{y_t}(x_t)$ are normalised and bounded away from $0$, 
\item every Markov kernel $\tau_{t,\theta}(dx_t|x_{t-1})$ has an associated pdf w.r.t. the Lebesgue measure, denoted $\tau_{t,\theta}^{x_{t-1}}(x_t)$, and both these pdf's and the likelihood functions $g_{t,\theta}^{y_t}(x_t)$ are Lipschitz continuous w.r.t. the parameter $\theta$. 
\end{enumerate}
These assumptions are made explicit in Section \ref{ssAssumptions}; then, in Sections \ref{ssUniformInner} and \ref{ssUniformOuter} we progress toward the main result in this paper, which can be outlined as follows.

\begin{Result} \label{res1}
(Theorem \ref{thUniform}, Section \ref{ssUniformOuter}).
If the assumptions (i), (ii) and (iii) above hold, and the jittering step of Algorithm \ref{alRA} is implemented using the kernel $\kappa_{N,\sfp}$ defined in \eqref{eqKappa}, then 
$$
\sup_{t\ge 0} \| (h,\mu_t^N) - (h,\mu_t) \|_p \le r(N)
$$ 
for every $h \in B(D_\theta)$ and $1 \le p \le \sfp$, where $r(N)$ is a rate function (to be given explicitly) such that $\lim_{N\rw\infty} r(N) = 0$. 
\end{Result}


Result \ref{res1} has some relevant consequences. In particular, in Section \ref{ssIdentification} we prove that, under the same regularity assumptions on the state-space model, it is possible to ``identify'' the static parameter $\Theta$, i.e., to compute estimates which are asymptotically exact.

\begin{Result} \label{res2}
(Theorem \ref{thIdentification}, Section \ref{ssIdentification}). 
If $\lim_{t\rw\infty} \mu_t = \delta_{\theta_*}$ for some $\theta_* \in D_\theta$, then 
$$
\limsup_{t\rw\infty} E\left[
	d(\mu_t^N,\delta_{\theta_*})
\right] \le \bar r(N),
$$
where 
\begin{itemize}
\item $d:\mP(D_\theta) \times \mP(D_\theta) \rw [0,+\infty)$ is a distance between probability measures, to be precisely defined in Section \ref{ssIdentification}, and
\item $\bar r(N)$ is a rate function (to be explicitly given) such that $\lim_{N\rw\infty} \bar r(N) = 0$.
\end{itemize}
\end{Result}

\section{Uniform convergence over time} \label{sUniform}

In this section we carry out the analysis leading to the uniform convergence over time of the approximation errors $\| (h,\mu_t^N) - (h,\mu_t) \|_p$, the explicit derivation of error rates and the asymptotically exact estimation of $\Theta$ (under regularity assumptions on the sequence $\{\mu_t\}_{t\ge0}$). Our argument is based on the approaches in \cite{DelMoral01c} and \cite{Kunsch05}, which rely on the stability of the sequences of measures to be approximated and the contractivity (under regularity assumptions) of the Markov kernels $\tau_{t,\tau}$. 

Within this setup, we show the uniform convergence of the particle filters in the inner layer (i.e., conditional on the value of the parameter) and then establish the same result for the complete Algorithm \ref{alRA}. This leads naturally to Result \ref{res2} on the asymptotically exact estimation of the static parameters. 

\subsection{Notation and assumptions} \label{ssAssumptions}

\subsubsection{Maps on the space of probability measures}


Recall that $\mP(\mX)$ and $\mP(D_\theta)$ denote the set of probability measures on $(\mX,\mB(\mX))$ and $(D_\theta,\mB(D_\theta))$, respectively. We introduce the map $\Psi_t^\theta : \mP(\mX) \rw \mP(\mX)$ that takes the predictive measure at time $t$ into the predictive measure at time $t+1$. A precise definition si given below.

\begin{Definicion}
For any integrable function $f:\mX\rw\Real$, any time $t\ge 0$ and any parameter vector $\theta\in D_\theta$, we define the map $\Psi_t^\theta : \mP(\mX) \rw \mP(\mX)$ as
\begin{equation}
\left(
	f,\Psi_t^\theta\left(
		\alpha
	\right)
\right) \dfn \frac{
	\left(
		g_{t-1,\theta}^{y_{t-1}}(f,\tau_{t,\theta}), \alpha
	\right)
}{
	\left(
		g_{t-1,\theta}^{y_{t-1}}, \alpha
	\right)
}.
\label{eqDefPsi}
\end{equation}
\end{Definicion}
It is simple to check (e.g., by way of Eqs. \eqref{eqPredicting} and \eqref{eqAlternative}) that $\xi_{t,\theta} = \Psi_t^\theta(\xi_{t-1,\theta})$ for $t \ge 2$. In order to define $\Psi_1^\theta$ in a consistent manner, let us introduce
\begin{eqnarray}
\phi_{-1} &\equiv& \mbox{the uniform measure on $\mX$, and}\nonumber\\ 
g_{0,\theta}^{y_0}(x) &=& g_0(x) \dfn 1 \quad \forall x \in \mX.\nonumber
\end{eqnarray}
Then, $\xi_0 = \phi_0 = \tau_0$ (independently of $\theta$) and $\xi_{1,\theta} = \Psi_1^\theta(\xi_0)$. Moreover, for any $0 \le k \le t$, let 
\begin{equation}
\Psi_{t|k}^\theta \dfn \Psi_t^\theta \circ \Psi_{t-1}^\theta \circ \cdots \circ \Psi_{k+1}^\theta,
\nonumber
\end{equation}
where $\circ$ denotes composition. Note that $\Psi_{t|t-1}^\theta = \Psi_t^\theta$ and we adopt the convention $\Psi_{t|t}^\theta(\alpha) = \alpha$.
 
\begin{Definicion} \label{defLambda}
For any integrable function $h:D_\theta\rw\Real$, any time $t>0$ and any $\alpha \in \mP(D_\theta)$, we define the map $\Lambda_t: \mP(D_\theta) \rw \mP(D_\theta)$ as 
\begin{equation}
\left(
	h, \Lambda_t(\alpha)
\right) \dfn \frac{
	(hu_t,\alpha)
}{
	(u_t,\alpha)
}, 
\nonumber
\end{equation}
hence $\mu_t=\Lambda_t(\mu_{t-1})$.
\end{Definicion}
The composition $\Lambda_{t|k} = \Lambda_t \circ \cdots \circ \Lambda_{k+1}$ is constructed in the same way as for $\Psi_{t|k}^\theta$. 

\subsubsection{Stability of the posterior probability measures}

Uniform convergence of particle filters over time can be guaranteed when the corresponding optimal filters satisfy some stability conditions \cite{DelMoral01c}. In a similar manner, here we adopt stability assumptions for the sequence of posterior probability measures (in $\mP(D_\theta)$) generated by the maps $\Lambda_t$, $t \ge 0$. These are made explicit below.
%
%

\begin{Assumption} \label{asStabilityLambda}
Let $\{ y_t \}_{t \ge 1}$ be an arbitrary sequence of observations  and let 
$$
\mS(h,T) = \sup_{\alpha,\eta \in \mP(D_\theta); k \ge 0} \left|
	\left(
		h, \Lambda_{k+T|k}(\alpha)
	\right) - \left(
		h, \Lambda_{k+T|k}(\eta)
	\right)
\right|,
$$
where $h : D_\theta \rw \Real$. Then, $\lim_{T\rw\infty} \mS(h,T) = 0$ for every $h \in B(D_\theta)$.
\end{Assumption}

\begin{Assumption} \label{asStabilityLambda-b}
For every $h \in B(D_\theta)$ there exist real constants $\bar b_1>0$ and $\bar b_2>0$, and a natural constant $\bar T_0 \ge 1$, such that 
$$
\mS(h,T) \le \bar b_1 \exp\left\{
	-\bar b_2 T
\right\} \quad \mbox{for every $T \ge \bar T_0$.}
$$
\end{Assumption}

\subsubsection{Bounds and Lipschitz continuity}

The latter stability assumptions for the maps $\Lambda_t$ are combined with the existence of certain bounds for the family of likelihood functions $g_{t,\theta}^{y_t}$ and Markov kernels $\tau_{t,\theta}$. These assumptions are made to ensure that the optimal inner filters (conditional on $\theta$) are stable for any choice of the parameters within the support $D_\theta$ \mbox{and} their particle approximations converge uniformly over time. They correspond to similar standard assumptions, e.g., in \cite{DelMoral04} or \cite{Kunsch05}, used in the analysis of conventional particle filters.
 
\begin{Assumption} \label{asBounds_g-1}
Let $\{ y_t \}_{t \ge 1}$ be an arbitrary but fixed sequence of observations. The likelihood functions are normalised and bounded away from 0, i.e., there exists a positive constant $a<\infty$ such that 
\begin{equation}
\inf_{x\in\mX, \theta\in D_\theta, t \ge 1} g_{t,\theta}^{y_t}(x) \ge \frac{1}{a}
\quad \mbox{and} \quad
\sup_{x\in\mX, \theta\in D_\theta, t \ge 1} g_{t,\theta}^{y_t}(x) \le 1.
\nonumber
\end{equation}
\end{Assumption}

Let $\tau_{t+m|t,\theta}(dx_{t+m}|x_t)$ denote the composition of $m$ consecutive Markov kernels, from time $t+1$ to time $t+m$, with starting point $x_t \in \mX$ at time $t$. In particular, the integral of a function $f:\mX\rw\Real$ w.r.t. the composite kernel  $\tau_{t+m|t,\theta}(dx_{t+m}|x_t)$ can be explicitly written as
\begin{equation}
(f,\tau_{t+m|t,\theta}(\cdot|x_t)) \dfn \int \cdots \int f(x_{t+m}) \tau_{t+m,\theta}(dx_{t+m}|x_{t+m-1}) \tau_{t+m-1,\theta}(dx_{t+m-1}|x_{t+m-2}) \cdots \tau_{t+1,\theta}(dx_{t+1}|x_t).
\nonumber 
\end{equation}
We make the following assumption on the composition of kernels.

\begin{Assumption} \label{asTau}
For a given integer $m >0$ there exists a constant $0 < \epsilon_\tau < 1$ such that, for every Borel set $A \in \mB(\mX)$,
\begin{equation}
\inf_{t \ge 0, (x,x')\in\mX^2, \theta\in D_\theta} \frac{
	\tau_{t+m|t,\theta}(A|x)
}{
	\tau_{t+m|t,\theta}(A|x')
} \ge \epsilon_\tau.
\nonumber
\end{equation}
\end{Assumption}

The jittering of the particles in the parameter space introduces a perturbation in the inner layer of particle filters of Algorithm \ref{alRA}. The procedure works when the effect of this perturbation on the approximating measures $\phi_{t,\theta}^N$ and $\xi_{t,\theta}^N$ is ``sufficiently small'', which can only be ensured when the corresponding measures enjoy some continuity property w.r.t. the parameters. This assumption is made explicit below.

\begin{Assumption} \label{asPsiLipschitz}
Every Markov kernel $\tau_{t,\theta}(dx|x')$ has a density w.r.t. the Lebesgue measure, denoted $\tau_{t,\theta}^{x'}(dx)$. The functions $g_{t,\theta}^{y_t}(x)$ and $\tau_{t,\theta}^{x'}(x)$ are Lipschitz in the parameter $\theta$ for every $(x,x') \in \mX^2$ and $t \ge 0$. In particular, there exists constants $L_g<\infty$ and $L_\tau$ such that, for any $\theta,\theta'\in D_\theta$,
\begin{eqnarray}
\sup_{t \ge 1; x \in \mX}| g_{t,\theta}^{y_t}(x) - g_{t,\theta'}^{y_t}(x) | &\le& L_g \| \theta - \theta' \|, \nonumber\\
\sup_{t\ge 0; (x,x')\in\mX^2} | \tau_{t,\theta}^{x'}(x) - \tau_{t,\theta'}^{x'}(x) | &\le& L_\tau \| \theta - \theta' \|. \nonumber
\end{eqnarray}
\end{Assumption}

\begin{Nota} \label{rmLipschitzs}
Let $L_{g,\tau} = L_g \vee L_\tau$. If assumptions A.\ref{asPsiLipschitz} and A.\ref{asBounds_g-1} hold, then it is not difficult to show that
\begin{equation}
\left|
	(f,\xi_{t,\theta})-(f,\xi_{t,\theta'})
\right| \le ta^t \| f \|_\infty L_{g,\tau} \| \theta - \theta' \|
\end{equation}
for any $f \in B(\mX)$ and $t \ge 1$, which corresponds to \cite[Assumption A.3]{Crisan15}. Integrals of the form $(f,\phi_{t,\theta})$ are also Lipschitz functions w.r.t. $\theta$, since 
$
(f,\xi_{t,\theta}) - (f,\xi_{t,\theta'}) = \left( 
	(f,\tau_{t,\theta}), \phi_{t-1,\theta} 
\right) - \left( 
	(f,\tau_{t,\theta'}), \phi_{t-1,\theta'} 
\right).
$
\end{Nota}

\subsubsection{An auxiliary result}

For any pair of integers $0<s<t$ we can explicitly construct the conditional pdf of the subsequence of observations $y_{s:t}$ given a point  $X_s=x_s$ in the state space and a choice parameters $\Theta=\theta$. We denote this density as $g_{s:t,\theta}^{y_{s:t}}(x_s)$, with the notation chosen to make explicit that, for fixed $y_{s:t}$, this is a function of the state value $x_s$ (i.e., it is interpreted as a likelihood). It is not difficult to show that
\begin{equation}
g_{s:t,\theta}^{y_{s:t}}(x_s) = \int \cdots \int \prod_{j=s}^t g_{j,\theta}^{y_j}(x_j) \prod_{l=s+1}^t \tau_{l,\theta}(dx_l|x_{l-1}).
\label{eqDefG}
\end{equation}

We also introduce a specific notation for the conditional distribution of the state $X_j$ conditional on $X_{j-1}=x_{j-1}$, $\Theta=\theta$ and the subsequence of observations from time $j$ up to time $t$, $y_{j:t}$. For any $j \le t$, this is a Markov kernel, denoted $\sfk_{j,\theta}^{y_{j:t}}(dx_j|x_{j-1})$, that can be explicitly written as 
\begin{equation}
\sfk_{j,\theta}^{y_{j:t}}(dx_j|x_{j-1}) = \frac{
	g_{j:t,\theta}^{y_{j:t}}(x_j) \tau_{j,\theta}(dx_j|x_{j-1})	
}{
	\int g_{j:t,\theta}^{y_{j:t}}(\tilde x_j) \tau_{j,\theta}(d \tilde x_j |x_{j-1})
}
\label{eqDefSfk}
\end{equation}
via the Bayes' theorem. If the observation sequence is fixed, then the composite probability measure 
\begin{equation}
\sfK_{s:t+1,\theta}^{y_{s:t}}(dx_{t+1}|x_s) = \int \cdots \int \tau_{t+1,\theta}(dx_{t+1}|x_t) \prod_{j=s+1}^t \sfk_{j,\theta}^{y_{j:t}}(dx_j|x_{j-1})
\label{eqComposK}
\end{equation}
is a Markov kernel on $(\mX,\mB(\mX))$.

The composite likelihood in \eqref{eqDefG} and the Markov kernel in \eqref{eqComposK} can be used to write integrals w.r.t. the composite map $\Psi_{t+1|s}^\theta$ explicitly. To be specific, given a probability measure $\alpha \in \mP(\mX)$, it is an exercise to show that
\begin{equation}
\left(
	f, \Psi_{t+1|s}^\theta(\alpha)
\right) = \frac{
	\left(
		(f, \sfK_{s:t+1,\theta}^{y_{s:t}}) g_{s:t,\theta}^{y_{s:t}}, \alpha
	\right)
}{
	\left(
		g_{s:t,\theta}^{y_{s:t}}, \alpha
	\right)
}.
\label{eqPsiNew}
\end{equation}
The representation in \eqref{eqPsiNew}, together with assumptions A.\ref{asBounds_g-1} and A.\ref{asTau}, enables the application of standard results from \cite{DelMoral04} which become instrumental in the analysis of Algorithm \ref{alRA}. 

We first define the Dobrushin contraction coefficient \cite{DelMoral01c} for Markov kernels and then show how it can be used to control the difference between between two probability measures $\Psi_{t+1|s}^\theta(\alpha)$ and $\Psi_{t+1|s}^\theta(\eta)$ which are constructed using the same composite map $\Psi_{t+1|s}^\theta$ (and, in particular, the same observation subsequence $y_{s:t+1}$) but different initial conditions $\alpha \ne \eta$.

\begin{Definicion}
The Dobrushin contraction coefficient of a Markov kernel $K_\theta$ from $\mX$ onto $(\mX,\mB(\mX))$ is 
$$
\beta(K_\theta) \dfn \sup_{x,x' \in \mX, A\in\mB(\mX)} \left|
	K_\theta(A|x) - K_\theta(A|x')
\right| \le 1.
$$
\end{Definicion}

An upper bound for the contraction coefficient of the kernel $\sfK_{s:t+1,\theta}^{y_{s:t}}$, explicitly given in terms of the constants $m$, $\epsilon_\tau$ and $a$ in assumptions A.\ref{asTau} and A.\ref{asBounds_g-1}, is given below.

\begin{Lema} \label{lmDobrushin}
If assumptions A.\ref{asBounds_g-1} and A.\ref{asTau} hold, then
\begin{equation}
\beta(\sfK_{s:t+1,\theta}^{y_{s:t}}) \le \left(
	1 - \frac{
		\epsilon_\tau^2
	}{
		a^{m-1}
	}
\right)^{ \lfloor \frac{t-s+1}{m} \rfloor }
\label{eqKcola0}
\end{equation}
for every $\theta \in D_\theta$.
\end{Lema}
\noindent {\bf Proof:} Since the inequalities in A.\ref{asBounds_g-1} and A.\ref{asTau} are assumed to hold uniformly over the parameter space $D_\theta$, the bound in \eqref{eqKcola0} follows readily from Proposition 4.3.3 in \cite{DelMoral04} (see also \cite[Corollary 4.3.3]{DelMoral04}). $\QED$

From Lemma \ref{lmDobrushin}, and given a test function $f \in B(\mX)$, we can obtain a bound for the difference $\left|  (f, \Psi_{t+1|s}^\theta(\alpha)) - (f, \Psi_{t+1|s}^\theta(\eta)) \right|$ that will ease considerably the convergence analysis for Algorithm \ref{alRA}. 

\begin{Lema} \label{lmDiffPsi}
Assume that A.\ref{asBounds_g-1} and A.\ref{asTau} hold true. Then, for any time indices $0 \le s \le t$, any pair of probability measures $\alpha, \eta \in \mP(\mX)$ and any test function $f \in B(\mX)$ there exists another bounded function $\tilde f_s \in B(\mX)$, with $\| f \|_\infty \le 1$, such that
\begin{equation}
\left|  
	(f, \Psi_{t+1|s}^\theta(\alpha)) - (f, \Psi_{t+1|s}^\theta(\eta)) 
\right| \le 2\| f \|_\infty \left(
	1 - \frac{
		\epsilon_\tau^2
	}{
		a^{m-1}
	}
\right)^{
	\lfloor
		\frac{
			t-s+1
		}{
			m
		}
	\rfloor
} \frac{
	a^m
}{
	\epsilon_\tau
} \left|
	(\tilde f_s,\alpha) - (\tilde f_s,\eta)
\right|.
\label{eqChula}
\end{equation}
\end{Lema}

\noindent{\bf Proof:} From \cite[Proposition 4.3.7]{DelMoral04} we obtain an upper bound for the difference of integrals that depends on the Dobrushin coefficient of the Markov kernel $\sfK_{s:t+1, \theta}^{y_{s:t}}$, namely
\begin{equation}
\left|  
	(f, \Psi_{t+1|s}^\theta(\alpha)) - (f, \Psi_{t+1|s}^\theta(\eta)) 
\right| \le 2\|f\|_\infty\beta(\sfK_{s:t+1, \theta}^{y_{s:t}}) \left(
	\sup_{x_s \in \mX} \frac{
		g_{s:t,\theta}^{y_{s:t}}(x_s)
	}{
		(g_{s:t,\theta}^{y_{s:t}},\alpha)
	}
\right) \left|
	(\tilde f_s, \alpha) - (\tilde f_s,\eta)
\right|,
\label{eqKcola1}
\end{equation}
for some $\tilde f_s : \mX \rw \Real$ with $\| \tilde f_s \| \le 1$. Moreover, from the definition of the composite likelihood in \eqref{eqDefG} and the assumption $g_{j,\theta}^{y_j} \le 1$ for every $j \ge 1$ and $\theta \in D_\theta$ (in A.\ref{asBounds_g-1}), it follows that
\begin{equation}
g_{s:t,\theta}^{y_{s:t}}(x_s) \le ( g_{s+m:t,\theta}^{y_{s+m:t}}, \tau_{s+m|s,\theta}(\cdot|x_s))
\label{eqKcola2}
\end{equation}
whereas, from the bound $g_{j,\theta}^{y_j}(x) \ge \frac{1}{a}$, for all $j \ge 1$ and $\theta \in D_{\theta}$ (in A.\ref{asBounds_g-1}) and the assumption A.\ref{asTau}, we obtain that
\begin{equation}
(g_{s:t,\theta}^{y_{s:t}},\alpha) \ge \frac{\epsilon_\tau}{a^m} ( g_{s+m:t}^{y_{s+m:t}}, \tau_{s+m|s,\theta}(\cdot|\tilde x_s))
\label{eqKcola3}
\end{equation}
for any $\tilde x_s \in \mX$. In particular, for $x_s=\tilde x_s$, the inequalities \eqref{eqKcola2} and \eqref{eqKcola3} taken together yield
$$
\frac{
	g_{s:t,\theta}^{y_{s:t}}(x_s)
}{
	(g_{s:t,\theta}^{y_{s:t}},\alpha)
} \le \frac{a^m}{\epsilon_\tau}
$$
independently of $x_s$. This, in turn, enables us to rewrite \eqref{eqKcola1} as 
\begin{equation}
\left|  
	(f, \Psi_{t+1|s}^\theta(\alpha)) - (f, \Psi_{t+1|s}^\theta(\eta)) 
\right| \le 2\|f\|_\infty\beta(\sfK_{s:t+1, \theta}^{y_{s:t}}) \frac{a^m}{\epsilon_\tau} \left|
	(\tilde f_s, \alpha) - (\tilde f_s,\eta)
\right|.
\label{eqKcola4}
\end{equation}
By combining Lemma \ref{lmDobrushin} with \eqref{eqKcola4} we readily obtain the inequality \eqref{eqChula} and complete the proof. $\QED$

\subsection{Uniform convergence of the inner particle filters} \label{ssUniformInner}

We first establish the uniform convergence over time of a conditional bootstrap filter when the parameter corresponds to a Markov chain with the kernel $\kappa_{N,\sfp}^{\theta'}(d\theta)$ described in Section \ref{ssJitteringKernel}. To be specific, assume that the model is the same as in Section \ref{ssModel} (in particular, the parameter $\Theta$ is random but fixed) however we run a modification of Algorithm \ref{alConditionalBF} where, at each time $t$, we generate a random variate $\theta_t$ with conditional probability measure $\kappa_{N,\sfp}^{\theta_{t-1}}(d\theta_t)$. The Markov chain is initialized with $\theta_0$ drawn from the prior $\mu_0$. The particle filter conditional on the chain $\{ \theta_t \}_{t\ge 0}$  constructed in this manner is outlined below. 

\begin{Algoritmo} \label{alMC-ConditionalBF}
Bootstrap filter conditional on a Markov chain of parameter realisations given by $\theta_0 \sim \mu_0(d\theta)$ and $\theta_t \sim \kappa_{N,\sfp}^{\theta_{t-1}}(d\theta)$, $t \ge 1$.
\begin{enumerate}
\item {\sf Initialisation.} Draw $N$ i.i.d. samples from $\tau_0$, denoted $x_0^{(i)}$, $i=1,\ldots,N$. 
\item {\sf Recursive step.} Let $\{ x_{t-1}^{(i)} \}_{1 \le i \le N}$ be the particles generated at time $t-1$. At time $t$, proceed with the two steps below.
        \begin{enumerate}
        \item For $i=1,...,N$, draw a sample $\bar x_t^{(i)}$ from the probability distribution $\tau_{t,\theta_t}(\cdot|x_{t-1}^{(i)})$ and compute the normalised weight
        \begin{equation}
        w_t^{(i)} = \frac{
                g_{t,\theta_t}^{y_{t}}(\bar x_t^{(i)})
        }{
                \sum_{k=1}^N g_{t,\theta_t}^{y_{t}}(\bar x_t^{(k)})
        }.
        \end{equation}
        
        \item For $i=1,...,N$, let $x_t^{(i)}=\bar x_t^{(k)}$ with probability $w_t^{(k)}$, $k \in \{1,...,N\}$.
        \end{enumerate}
\end{enumerate}
\end{Algoritmo}

Note that, for any particle $\bar \theta_t^{(i)}$, $i \in \{1, ..., N\}$, at time $t$ in the nested particle filter described by Algorithm \ref{alRA}, each conditional particle filter in the inner layer can be described as an instance of Algorithm \ref{alMC-ConditionalBF}. Indeed, by tracking the ``history'' of $\bar \theta_t^{(i)}$ across the resampling steps of Algorithm \ref{alRA}, we find that there is a sequence on $D_\theta$ of the form
$
\theta_{0|t}^{(i)}, \theta_{1|t}^{(i)}, \ldots, \theta_{t|t}^{(i)}
$
such that, 
\begin{itemize}
\item for $n=0$, $\theta_{0|t}^{(i)}$ is drawn from $\mu_0$,
\item for any $0 \le n\le t$, $\theta_{n|t}^{(i)}$ is drawn from the kernel $\kappa_{N,\sfp}^{\theta_{n-1|t}^{(i)}}$ and, 
\item for $n=t$, $\theta_{t|t}^{(i)} = \bar \theta_t^{(i)}$. 
\end{itemize}
Lemma \ref{lmUniformPhi} below states that the approximation $(f,\xi_{t,\theta_t}^N)$, where $f \in B(\mX)$, generated by  Algorithm \ref{alMC-ConditionalBF} actually converges to $(f,\xi_{t,\theta_t})$, as $N$ increases, uniformly over time under a subset of the assumptions in Section \ref{ssAssumptions}. This is a non-trivial result. Note that $\xi_{t,\theta_t}$ is the predictive probability measure at time $t$ associated to the state space model $\{ \tau_0, \tau_{n,\Theta}, g_{n,\Theta}^{y_n} \}_{1 \le n \le t}$, where $\Theta=\theta_t$ is fixed, while $\xi_{t,\theta_t}^N$ results from Algorithm \ref{alMC-ConditionalBF}, where the parameter value is effectively changing over time as a realisation $\theta_0, \theta_1, ..., \theta_t$ of a Markov chain up to time $t$.

\begin{Lema} \label{lmUniformPhi}
Let $\{ \theta_t \}_{t\ge 0}$ denote a Markov chain on the compact set $D_\theta$, generated from the prior $\mu_0$ and the kernels $\kappa_{N,\sfp}^{\theta_{t-1}}(d\theta)$ constructed as in Eq. \eqref{eqKappa}. Let $\xi_{t,\theta_t}^N = \frac{1}{N}\sum_{n=1}^N \delta_{\bar x_t^{(n)}}$ be the sequence of approximate predictive measures generated by Algorithm \ref{alMC-ConditionalBF}. If assumptions A.\ref{asBounds_g-1}, A.\ref{asTau} and A.\ref{asPsiLipschitz} hold then there exists a real constant $\bar C$, independent of $N$ and independent of the sequence $\{ \theta_t \}_{t \ge 0}$, such that, for any $f \in B(\mX)$ and any $1 \le p \le \sfp$,
\begin{equation}
\sup_{t \ge 0} \left\|
	(f,\xi_{t,\theta_t}^N) - (f,\xi_{t,\theta_t})
\right\|_p \le \frac{
	\bar C
}{
	\sqrt{N}
}.
\label{eqBoundInnerLayer}
\end{equation}
In particular, 
$\lim_{N\rw\infty} \sup_{t \ge 0} \|
	(f,\xi_{t,\theta_t}^N) - (f,\xi_{t,\theta_t})
\|_p = 0$.
\end{Lema}

\noindent \textbf{\textit{Proof:}} We look into the approximation error $\left| (f,\xi_{t,\theta_t}^N) - (f,\xi_{t,\theta_t}) \right|$, which can be written as
\begin{eqnarray}
\left|
	(f,\xi_{t,\theta_t}^N) - (f,\xi_{t,\theta_t}) 
\right| &=& \left|
	\sum_{k=0}^{t-1} \left(
		f, \Psi_{t|t-k}^{\theta_t}\left(
			\xi_{t-k,\theta_{t-k}}^N
		\right)
	\right) - \left(
		f, \Psi_{t|t-k-1}^{\theta_t}\left(
			\xi_{t-k-1,\theta_{t-k-1}}^N
		\right)
	\right) 
\right. \nonumber \\
&& \left.
	+ \left(
		f, \Psi_{t|0}^{\theta_t}\left(
			\xi_{0,\theta_0}^N
		\right)
	\right) - \left(
		f, \Psi_{t|0}^{\theta_t}\left(
			\tau_0
		\right)
	\right)
\right| \nonumber\\
&\le& \sum_{k=0}^{t-1} \left|
	\left(
		f, \Psi_{t|t-k}^{\theta_t}\left(
			\xi_{t-k,\theta_{t-k}}^N
		\right)
	\right) - \left(
		f, \Psi_{t|t-k-1}^{\theta_t}\left(
			\xi_{t-k-1,\theta_{t-k-1}}^N
		\right)
	\right)
\right| \nonumber\\
&& + \left|
	\left(
		f, \Psi_{t|0}^{\theta_t}\left(
			\xi_{0,\theta_0}^N
		\right)
	\right) - \left(
		f, \Psi_{t|0}^{\theta_t}\left(
			\tau_0
		\right)
	\right)
\right|,
\label{eqDecompo-Psi0}
\end{eqnarray}
where the equality follows from a `telescopic' decomposition of the difference $(f,\xi_{t,\theta_t}^N) - (f,\xi_{t,\theta_t})$. To see this, simply recall that $\xi_{0,\theta_0}^N \equiv \phi_{0,\theta_0}^N \equiv \tau_0^N$ (independently of $\theta_0$ according to the model in Section \ref{ssModel}) and note that $\Psi_{t|0}^{\theta_t}(\tau_0) = \xi_{t,\theta_t}$. By way of Minkowski's inequality, \eqref{eqDecompo-Psi0} enables us to express the $L_p$ norm of the approximation error (for $p \ge 1$) as
\begin{eqnarray}
\left\|
	(f,\xi_{t,\theta_t}^N) - (f,\xi_{t,\theta_t}) 
\right\|_p &\le& \sum_{k=0}^{t-1} \left\|
	\left(
		f, \Psi_{t|t-k}^{\theta_t}\left(
			\xi_{t-k,\theta_{t-k}}^N
		\right)
	\right) - \left(
		f, \Psi_{t|t-k-1}^{\theta_t}\left(
			\xi_{t-k-1,\theta_{t-k-1}}^N
		\right)
	\right)
\right\|_p \nonumber\\
&& + \left\|
	\left(
		f, \Psi_{t|0}^{\theta_t}\left(
			\xi_{0,\theta_0}^N
		\right)
	\right) - \left(
		f, \Psi_{t|0}^{\theta_t}\left(
			\tau_0
		\right)
	\right)
\right\|_p,
\label{eqDecompo-Psi}
\end{eqnarray}

The last term in the decomposition above can be easily upper bounded using Lemma \ref{lmDiffPsi}, namely
\begin{eqnarray}
\left\|
	\left(
		f, \Psi_{t|0}^{\theta_t}\left(
			\xi_{0,\theta_0}^N
		\right)
	\right) - \left(
		f, \Psi_{t|0}^{\theta_t}\left(
			\tau_0
		\right)
	\right)
\right\|_p &\le& 2\| f \|_\infty \left(
	1 - \frac{
		\epsilon_\tau^2
	}{
		a^{m-1}
	}
\right)^{
	\lfloor
		\frac{
			t
		}{
			m
		}
	\rfloor
} \frac{
	a^m
}{
	\epsilon_\tau
} \left\|
	(\tilde f_0,\tau_0^N) - (\tilde f_0,\tau_0)
\right\|_p, \nonumber \\
&\le& 2\| f \|_\infty \left(
	1 - \frac{
		\epsilon_\tau^2
	}{
		a^{m-1}
	}
\right)^{
	\lfloor
		\frac{
			t
		}{
			m
		}
	\rfloor
} \frac{
	a^m
}{
	\epsilon_\tau
} \frac{\tilde C_0}{\sqrt{N}}
\label{eqSup_mE}
\end{eqnarray}
where $\| \tilde f_0 \|_\infty \le 1$ and the second inequality follows readily from the fact that $\tau_0^N = \xi_{0,\theta_0}^N$ is an i.i.d. Monte Carlo approximation of $\tau_0$ (hence, $\tilde C_0<\infty$ is a constant independent of $N$). For the remaining terms in the sum of \eqref{eqDecompo-Psi}, Lemma \ref{lmDiffPsi} yields
\begin{eqnarray}
\left\|
	\left(
		f, \Psi_{t|t-k}^{\theta_t}\left(
			\xi_{t-k,\theta_{t-k}}^N
		\right)
	\right) - \left(
		f, \Psi_{t|t-k-1}^{\theta_t}\left(
			\xi_{t-k-1,\theta_{t-k-1}}^N
		\right)
	\right)
\right\|_p &\le& \nonumber\\
2\| f \|_\infty \left(
	1 - \frac{
		\epsilon_\tau^2
	}{
		a^{m-1}
	}
\right)^{
	\lfloor
		\frac{
			k
		}{
			m
		}
	\rfloor
} \frac{
	a^m
}{
	\epsilon_\tau
} \left\|
	\left( \tilde f_{t-k},\xi_{t-k,\theta_{t-k}}^N \right) - \left( \tilde f_{t-k},\Psi_{t-k}^{\theta_t}\left(
			\xi_{t-k-1,\theta_{t-k-1}}^N
		\right) \right)
\right\|_p.&&
\label{eqKcurra0}
\end{eqnarray}
where $\| \tilde f_{t-k} \|_\infty \le 1$.

In order to convert \eqref{eqKcurra0} into an explicit error rate, we need to derive bounds for errors of the form
$
\left\|
	\left(
		h, \xi_{t-k,\theta_{t-k}}^N
	\right) - \left(
		h, \Psi_{t-k}^{\theta_t}\left(
			\xi_{t-k-1,\theta_{t-k-1}}^N
		\right)
	\right)
\right\|_p
$, where $h:\mX \rw \Real$ with $\| h \|_\infty \le 1$. With this aim, we consider the triangular inequality
\begin{eqnarray}
\left\|
	\left(
		h, \xi_{t-k,\theta_{t-k}}^N
	\right) - \left(
		h, \Psi_{t-k}^{\theta_t}\left(
			\xi_{t-k-1,\theta_{t-k-1}}^N
		\right)
	\right)
\right\|_p 
\le
\left\|
	\left(
		h, \xi_{t-k,\theta_{t-k}}^N
	\right) - E\left[
		\left(
			h, \xi_{t-k,\theta_{t-k}}^N
		\right) | \mG_{t-k}
	\right]
\right\|_p + \nonumber\\
\left\|
	E\left[
		\left(
			h, \xi_{t-k,\theta_{t-k}}^N
		\right) | \mG_{t-k}
	\right] - \left(
		h, \Psi_{t-k}^{\theta_t}\left(
			\xi_{t-k-1,\theta_{t-k-1}}^N
		\right)
	\right)
\right\|_p, &&\label{eqChori8}
\end{eqnarray}
where
$
\mG_{t-k} = \sigma\left(
	x_{0:t-k-1}^{(n)}, \bar x_{1:t-k-1}^{(n)}, \{ \theta_s \}_{s\ge 0}; 1 \le n \le N
\right)
$
is the $\sigma$-algebra generated by the random variables between brackets, and analyse the two terms on the right hand side separately.

For the first term on the right hand side of \eqref{eqChori8}, we note that 
\begin{equation}
\left(
	h, \xi_{t-k,\theta_{t-k}}^N
\right) - E\left[
	\left(
		h, \xi_{t-k,\theta_{t-k}}^N
	\right) | \mG_{t-k}
\right] = \frac{1}{N} \sum_{n=1}^N \bar S_{t-k}^{(n)},
\nonumber
\end{equation}
where 
\begin{equation}
\bar S_{t-k}^{(n)} = h(\bar x_{t-k}^{(n)}) - E\left[
	h(\bar x_{t-k}^{(n)}) | \mG_{t-k}
\right], \quad n=1, ..., N,
\nonumber
\end{equation}
are zero-mean and conditionally (on $\mG_{t-k}$) independent r.v.'s. Therefore it is straightforward to show that  
\begin{equation}
E\left[
	\left|
		\left(
			h, \xi_{t-k,\theta_{t-k}}^N
		\right) - E\left[
			\left(
				h, \xi_{t-k,\theta_{t-k}}^N
			\right) | \mG_{t-k}
		\right]
	\right|^p \left| \mG_{t-k} \right.
\right] = E\left[
	\left|
		\frac{1}{N} \sum_{n=1}^N \bar S_{t-k}^{(n)}
	\right|^p \left| \mG_{t-k} \right.
\right] \le \frac{
	c^p
}{
 	N^\frac{p}{2}
}
\label{eqChori9}
\end{equation}
for some constant $c>0$ independent of $N$ and independent of the distribution of the variables $\bar S_{t-k}^{(n)}$, $n=1, ..., N$ (in particular, independent of the sequence $\{\theta_t\}_{t\ge 0}$). Taking expectations on both sides of \eqref{eqChori9}, and then exponentiating by $\frac{1}{p}$, yields
\begin{equation}
\left\|
	\left(
		h, \xi_{t-k,\theta_{t-k}}^N
	\right) - E\left[
		\left(
			h, \xi_{t-k,\theta_{t-k}}^N
		\right) | \mG_{t-k}
	\right]
\right\|_p \le \frac{
	c
}{
 	\sqrt{N}
}.
\label{eqChori10}
\end{equation}

To find a rate for the second term in \eqref{eqChori8}, we note that
\begin{equation}
E\left[
	\left(
		h, \xi_{t-k,\theta_{t-k}}^N
	\right) | \mG_{t-k}
\right] = \frac{
	\left(
		g_{t-k-1,\theta_{t-k-1}}^{y_{t-k-1}} \left(
			h, \tau_{t-k,\theta_{t-k}}
		\right), \xi_{t-k-1,\theta_{t-k-1}}^N
	\right)
}{
	\left(
		g_{t-k-1,\theta_{t-k-1}}^{y_{t-k-1}}, \xi_{t-k-1,\theta_{t-k-1}}^N
	\right)
}
\label{eqChori11}
\end{equation}
whereas
\begin{equation}
\left(
	h, \Psi_{t-k}^{\theta_t}\left(
		\xi_{t-k-1,\theta_{t-k-1}}^N
	\right)
\right) = \frac{
	\left(
		g_{t-k-1,\theta_t}^{y_{t-k-1}} \left(
			h, \tau_{t-k,\theta_t}
		\right), \xi_{t-k-1,\theta_{t-k-1}}^N
	\right)
}{
	\left(
		g_{t-k-1,\theta_t}^{y_{t-k-1}}, \xi_{t-k-1,\theta_{t-k-1}}^N
	\right)
}.
\label{eqChori12}
\end{equation}
Subtracting \eqref{eqChori12} from \eqref{eqChori11} and then rearranging terms yields
\begin{eqnarray}
E\left[
	\left(
		h, \xi_{t-k,\theta_{t-k}}^N
	\right) | \mG_{t-k}
\right] - \left(
	h, \Psi_{t-k}^{\theta_t}\left(
		\xi_{t-k-1,\theta_{t-k-1}}^N
	\right)
\right) &=& \nonumber\\
\frac{
	\left(
		g_{t-k-1,\theta_{t-k-1}}^{y_{t-k-1}} \left(
			h, \tau_{t-k,\theta_{t-k}}
		\right) - g_{t-k-1,\theta_t}^{y_{t-k-1}} \left(
			h, \tau_{t-k,\theta_t}
		\right), \xi_{t-k-1,\theta_{t-k-1}}^N
	\right)
}{
	\left(
		g_{t-k-1,\theta_t}^{y_{t-k-1}}, \xi_{t-k-1,\theta_{t-k-1}}^N
	\right)
} &+& \nonumber \\
\frac{
	E\left[
		\left(
			h, \xi_{t-k,\theta_{t-k}}^N
		\right) | \mG_{t-k}
	\right] \times \left(
		g_{t-k-1,\theta_t}^{y_{t-k-1}} - g_{t-k-1,\theta_{t-k-1}}^{y_{t-k-1}}, 
		\xi_{t-k-1,\theta_{t-k-1}}^N
	\right) 
}{
	\left(
		g_{t-k-1,\theta_t}^{y_{t-k-1}}, \xi_{t-k-1,\theta_{t-k-1}}^N
	\right)	
}, && \nonumber
\end{eqnarray}
hence
\begin{eqnarray}
\left|
	E\left[
		\left(
			h, \xi_{t-k,\theta_{t-k}}^N
		\right) | \mG_{t-k}
	\right] - \left(
		h, \Psi_{t-k}^{\theta_t}\left(
			\xi_{t-k-1,\theta_{t-k-1}}^N
		\right)
	\right) 
\right| &\le& \nonumber\\
a\times \left(
	\left|
		g_{t-k-1,\theta_{t-k-1}}^{y_{t-k-1}} \left(
			h, \tau_{t-k,\theta_{t-k}}
		\right) - g_{t-k-1,\theta_t}^{y_{t-k-1}} \left(
			h, \tau_{t-k,\theta_t}
		\right)
	\right|, \xi_{t-k-1,\theta_{t-k-1}}^N
\right) &+& \nonumber\\
a \times \left(
	\left|
		g_{t-k-1,\theta_t}^{y_{t-k-1}} - g_{t-k-1,\theta_{t-k-1}}^{y_{t-k-1}}
	\right|, \xi_{t-k-1,\theta_{t-k-1}}^N
\right), && 
\label{eqChori13}
\end{eqnarray} 
where we have used the obvious bounds
$
E\left[
	\left(
		h, \xi_{t-k,\theta_{t-k}}^N
	\right) | \mG_{t-k}
\right] \le \| h \|_\infty \le 1
$ and, from assumption A.\ref{asBounds_g-1},
$
\left(
	g_{t-k-1,\theta_t}^{y_{t-k-1}}, \xi_{t-k-1,\theta_{t-k-1}}^N
\right) \ge a^{-1}.
$ 

From assumption A.\ref{asPsiLipschitz}, the likelihoods $g_{t,\theta}^{y_t}(x)$ are Lipschitz in the parameter $\theta$, with constant $L_g$ independent of $t$ and $x$. In particular, 
\begin{equation}
\sup_{x \in \mX, t \ge T} \left|
	g_{t-k-1,\theta_t}^{y_{t-k-1}}(x) - g_{t-k-1,\theta_{t-k-1}}^{y_{t-k-1}}(x)
\right| \le L_g\| \theta_t - \theta_{t-k-1} \|.
\label{eqLips-1}
\end{equation}
Also from assumption A.\ref{asPsiLipschitz}, the kernels $\tau_{t,\theta}(dx|x) \in \mP(\mX)$ are endowed with densities w.r.t. the Lebesgue measure, hence we can write
\begin{eqnarray}
\left|
	g_{t-k-1,\theta_{t-k-1}}^{y_{t-k-1}}(x) \left(
		h, \tau_{t-k,\theta_{t-k}}
	\right)(x) - g_{t-k-1,\theta_t}^{y_{t-k-1}}(x) \left(
		h, \tau_{t-k,\theta_t}
	\right)(x)
\right| &=& \nonumber \\
\left|
	g_{t-k-1,\theta_{t-k-1}}^{y_{t-k-1}}(x) 
	\int h(x') \tau_{t-k,\theta_{t-k}}^x(x') dx'
	- g_{t-k-1,\theta_t}^{y_{t-k-1}}(x) 
	 \int h(x') \tau_{t-k,\theta_t}^x(x') dx'
\right| \nonumber
\end{eqnarray}
and a simple triangle inequality yields
\begin{eqnarray}
\left|
	g_{t-k-1,\theta_{t-k-1}}^{y_{t-k-1}}(x) \left(
		h, \tau_{t-k,\theta_{t-k}}
	\right)(x) - g_{t-k-1,\theta_t}^{y_{t-k-1}}(x) \left(
		h, \tau_{t-k,\theta_t}
	\right)(x)
\right| &\le& \nonumber \\
\left|
	\left(
		g_{t-k-1,\theta_{t-k-1}}^{y_{t-k-1}}(x) - 
		g_{t-k-1,\theta_{t-k}}^{y_{t-k-1}}(x)
	\right)  \int h(x') \tau_{t-k,\theta_{t-k}}^x(x')dx'
\right| + &&\nonumber\\
\left|
	\int h(x') \left(
		g_{t-k-1,\theta_{t-k}}^{y_{t-k-1}}(x)\tau_{t-k,\theta_{t-k}}^x(x') -
		g_{t-k-1,\theta_t}^{y_{t-k-1}}(x)\tau_{t-k,\theta_t}^x(x')
	\right) dx'
\right| &\le& \nonumber\\
L_g \vee L_{g,\tau} \left(
	\| \theta_{t-k-1} - \theta_{t-k} \| + \| \theta_t - \theta_{t-k} \|
\right), \label{eqLips-2}
\end{eqnarray}
where the second inequality is satisfied because the product $g_{t,\theta}^{y_t}\tau_{t,\theta'}^x(x')$ is Lipschitz in $\theta$ for every $t \ge 1$ and $x,x' \in \mX$ (a consequence of assumption A.\ref{asPsiLipschitz}) with constant $L_{g,\tau}$.

If we substitute \eqref{eqLips-1} and \eqref{eqLips-2} back into \eqref{eqChori13} we obtain
\begin{equation}
\left|
	E\left[
		\left(
			h, \xi_{t-k,\theta_{t-k}}^N
		\right) | \mG_{t-k}
	\right] - \left(
		h, \Psi_{t-k}^{\theta_t}\left(
			\xi_{t-k-1,\theta_{t-k-1}}^N
		\right)
	\right) 
\right| \le 2a L \sum_{j=0}^{k} \| \theta_{t-j} - \theta_{t-j-1} \|
\label{eqChori14}
\end{equation} 
where we have introduced the constant $L=\max\{L_g,L_{g,\tau}\}$ and taken advantage of the straightforward inequality
$
\| \theta_t - \theta_{t-k-1} \| \le \sum_{j=0}^{k} \| \theta_{t-j} - \theta_{t-j-1} \|.
$ 
Raising both sides of \eqref{eqChori14} to power $p$ and then taking expectations yields
\begin{eqnarray}
E\left[
	\left|
		E\left[
			\left(
				h, \xi_{t-k,\theta_{t-k}}^N
			\right) | \mG_{t-k}
		\right] - \left(
			h, \Psi_{t-k}^{\theta_t}\left(
				\xi_{t-k-1,\theta_{t-k-1}}^N
			\right)
		\right) 
	\right|^p
\right] &\le& (	2aL )^p E\left[
	\left|
		\sum_{j=0}^{k} \| \theta_{t-j} - \theta_{t-j-1} \|
	\right|^p
\right] \nonumber\\
&\le& \left( 2a L (k+1) \right)^p \times \nonumber\\
&& \times \frac{1}{k+1}\sum_{j=0}^k E\left[
	\| \theta_{t-j} - \theta_{t-j-1} \|^p
\right], 
\label{eqChori15}
\end{eqnarray} 
where \eqref{eqChori15} follows from Jensen's inequality. Combining \eqref{eqChori15} with Proposition \ref{propAssumptionKappa1} 
we arrive at
\begin{equation}
\left\|
	E\left[
		\left(
			h, \xi_{t-k,\theta_{t-k}}^N
		\right) | \mG_{t-k}
	\right] - \left(
		h, \Psi_{t-k}^{\theta_t}\left(
			\xi_{t-k-1,\theta_{t-k-1}}^N
		\right)
	\right) 
\right\|_p \le 2aL (k+1) \frac{
	c_\kappa
}{
	\sqrt{N}
},
\label{eqChori15.5}
\end{equation}
where $c_\kappa < \infty$ is a constant independent of $N$, $t$ and $\{ \theta_n \}_{n\ge 0}$. 

If we now insert \eqref{eqChori10} and \eqref{eqChori15.5} into \eqref{eqChori8} we obtain the relationship 
\begin{equation}
\left\|
	\left(
		h, \xi_{t-k,\theta_{t-k}}^N
	\right) - \left(
		h, \Psi_{t-k}^{\theta_t}\left(
			\xi_{t-k-1,\theta_{t-k-1}}^N
		\right)
	\right)
\right\|_p 
\le
\frac{
	c + 2aL(k+1)c_\kappa 
}{
 	\sqrt{N}
}, \label{eqChori16}
\end{equation}
where the numerator is finite and constant w.r.t. $N$, $\{ \theta_n \}_{n\ge 0}$ and $t$. At this point, we only need to substitute the latter inequality backwards. Indeed, if we plug \eqref{eqChori16}, with $h = \tilde f_{t-k}$, into \eqref{eqKcurra0} and then substitute the resulting bound, together with \eqref{eqSup_mE}, into \eqref{eqDecompo-Psi}, we arrive at
\begin{equation}
\left\|
	(f,\xi_{t,\theta_t}^N) - (f,\xi_{t,\theta_t}) 
\right\|_p \le \frac{
	2\| f \|_\infty a^m \epsilon_\tau^{-1}
}{
	\sqrt{N}
} \sum_{k=0}^t 
\left(
	1 - \frac{
		\epsilon_\tau^2
	}{
		a^{m-1}
	}
\right)^{
	\lfloor
		\frac{
			k
		}{
			m
		}
	\rfloor
}(\bar C_0 + \bar C_1 k),
\label{eqKcurra1}
\end{equation}
where $\bar C_0 = c+2aLc_\kappa$ and $\bar C_1 = \tilde C_0 \vee 2aLc_\kappa$. 

What remains to be proved is that the sum in \eqref{eqKcurra1} admits an upper bound $\bar C < \infty$ independent of $t$. To show this, we decompose 
\begin{equation}
\sum_{k=0}^t 
\left(
	1 - \frac{
		\epsilon_\tau^2
	}{
		a^{m-1}
	}
\right)^{
	\lfloor
		\frac{
			k
		}{
			m
		}
	\rfloor
}(\bar C_0 + \bar C_1 k) = \bar C_0 \sum_{k=0}^t 
\left(
	1 - \frac{
		\epsilon_\tau^2
	}{
		a^{m-1}
	}
\right)^{
	\lfloor
		\frac{
			k
		}{
			m
		}
	\rfloor
} + \bar C_1 \sum_{k=0}^t 
k \left(
	1 - \frac{
		\epsilon_\tau^2
	}{
		a^{m-1}
	}
\right)^{
	\lfloor
		\frac{
			k
		}{
			m
		}
	\rfloor
}
\label{eqKcurra2}
\end{equation}
and note that each term in \eqref{eqKcurra2} can be written as a sum of convergent series. Indeed, for the first term we have 
\begin{eqnarray}
\sum_{k=0}^t \left(
	1 - \frac{
		\epsilon_\tau^2
	}{
		a^{m-1}
	}
\right)^{
	\lfloor
		\frac{
			k
		}{
			m
		}
	\rfloor
} &\le& m \sum_{k=0}^\infty
\left(
	1 - \frac{
		\epsilon_\tau^2
	}{
		a^{m-1}
	}
\right)^k \label{eqKcurra4} \\
&=& m a^{m-1} \epsilon_\tau^{-2},
\label{eqKcurra5}
\end{eqnarray}
where the inequality \eqref{eqKcurra4} is obtained from the identity $\sum_{k=0}^\infty r^{\lfloor \frac{k}{m} \rfloor} = m \sum_{k=0}^\infty r^k$ (for any $r \in (0,1)$) and \eqref{eqKcurra5}  follows from the limit of the geometric series. For the second term in \eqref{eqKcurra2} we have 
\begin{eqnarray}
\sum_{k=0}^t k \left(
	1 - \frac{
		\epsilon_\tau^2
	}{
		a^{m-1}
	}
\right)^{
	\lfloor
		\frac{
			k
		}{
			m
		}
	\rfloor
} &\le& 2m \sum_{k=0}^\infty
\left\lfloor \frac{k}{m} \right\rfloor \left(
	1 - \frac{
		\epsilon_\tau^2
	}{
		a^{m-1}
	}
\right)^{\left\lfloor \frac{k}{m} \right\rfloor}\label{eqKcurra6} \\
&=& 2m^2 \sum_{k=0}^\infty k \left(
	1 - \frac{
		\epsilon_\tau^2
	}{
		a^{m-1}
	}
\right)^k,
\label{eqKcurra7} \\
&=& 2m^2 \frac{
	1 - \epsilon_\tau^2 a^{-(m-1)}
}{
	\epsilon_\tau^2 a^{-2(m-1)}
}, \label{eqKcurra8}
\end{eqnarray}
where \eqref{eqKcurra6} follows from the inequality $k \le 2m\lfloor \frac{k}{m} \rfloor$ (for $k =0, 1, 2, ...$ and $m \ge 1$), \eqref{eqKcurra7} holds because of the identity $\sum_{k=0}^\infty \lfloor \frac{k}{m} \rfloor r^{\lfloor \frac{k}{m} \rfloor} = m \sum_{k=0}^\infty k r^k$ (for any $r\in(0,1)$) and \eqref{eqKcurra8} is readily obtained from the limit $\sum_{k=0}^\infty kr^k = \frac{r}{(1-r)^2}$ (for $|r|<1$).

To conclude the proof, we simply put \eqref{eqKcurra1}, \eqref{eqKcurra2}, \eqref{eqKcurra5} and \eqref{eqKcurra8} together, to obtain the desired inequality \eqref{eqBoundInnerLayer} with 
\begin{equation}
\bar C = 2\| f \|_\infty a^m \epsilon_\tau^{-1}\left(
	\bar C_0 m a^{m-1} \epsilon_\tau^{-2} + 2 \bar C_1 m^2 \frac{
	1 - \epsilon_\tau^2 a^{-(m-1)}
}{
	\epsilon_\tau^2 a^{-2(m-1)}
}
\right) \le 4\| f \|_\infty (\bar C_0 \vee \bar C_1) \epsilon_\tau^{-3} a^{3m}
\label{eqLaUltima}
\end{equation}
and $\bar C_0 \vee \bar C_1 \le a( c + \tilde C_0 + 2Lc_\kappa )$.
$\QED$

\subsection{Uniform convergence of the nested particle filter} \label{ssUniformOuter}

Lemma \ref{lmUniformPhi} can be used to obtain bounds for the errors in the computation of the weights of Algorithm \ref{alRA}. Based on this result, it is possible to show that the overall procedure converges uniformly over time given the assumptions in Section \ref{ssUniformInner}, and provide an error rate. This is explicitly given by the following theorem.

\begin{Teorema} \label{thUniform}
Let $\{ y_t \}_{t \ge 1}$ be an arbitrary sequence of observations, let  $D_\theta$ be a compact set and select a jittering kernel $\kappa_{N,\sfp}$ from the family in Eq. \eqref{eqKappa}. If assumptions A.\ref{asStabilityLambda}, A.\ref{asBounds_g-1}, A.\ref{asTau} and A.\ref{asPsiLipschitz} are satisfied, 
then 
\begin{equation}
\lim_{N\rw\infty} \sup_{t\ge 0} \| (h,\mu_t^N) - (h,\mu_t) \|_p = 0
\nonumber
\end{equation}
for any $h \in B(D_\theta)$ and $1 \le p \le \sfp$. If, additionally, the exponential stability assumption A.\ref{asStabilityLambda-b} holds, then there exists $C<\infty$, independent of $N$ and $t$, such that
\begin{equation}
\sup_{t \ge 0} \left\|
	(h,\mu_t^N)-(h,\mu_t)
\right\|_p \le N^{-\frac{1}{2} + \epsilon} + C N^{-\epsilon\frac{\bar b_2}{1+\log(a)}}
\nonumber
\end{equation}
for any $0 < \epsilon < \frac{1}{2}$, where $C<\infty$ is a constant independent of $N$ and $t$, while $a$ and $\bar b_2$ are the constants specified in assumptions A.\ref{asBounds_g-1} and A.\ref{asStabilityLambda-b}.
\end{Teorema}

\noindent \textit{\textbf{Proof:}} Choose some integer $T>0$. We look into the error $\| (h,\mu_t^N)-(h,\mu_t) \|_p$ for $t<T$ and $t\ge T$ separately. 

For any $t \ge T$, the difference $(h,\mu_t^N)-(h,\mu_t)$ can be decomposed as
\begin{eqnarray}
(h,\mu_t^N)-(h,\mu_t) &=& \sum_{k=0}^{T-1} \left(
	h, \Lambda_{t|t-k}(\mu_{t-k}^N)
\right) - \left( 
	h, \Lambda_{t|t-k-1}(\mu_{t-k-1}^N)
\right) \nonumber \\
&&+ \left(
	h, \Lambda_{t|t-T}(\mu_{t-T}^N)
\right) - \left(
	h, \Lambda_{t|t-T}(\mu_{t-T})
\right). \label{eqGordo1}
\end{eqnarray}
The last term on the right hand side of \eqref{eqGordo1} can be bounded using A.\ref{asStabilityLambda},  namely 
\begin{equation}
\left|
	\left(
		h, \Lambda_{t|t-T}(\mu_{t-T}^N)
	\right) - \left(
		h, \Lambda_{t|t-T}(\mu_{t-T})
	\right)
\right| \le \mS(h,T),
\label{eqGordo1.5}
\end{equation}
where $\mS(h,T)$ is independent of $N$ and $t$, and $\lim_{T\rw\infty} \mS(h,T)=0$ for every $h \in B(D_\theta)$. Minkowski's inequality, together with \eqref{eqGordo1} and \eqref{eqGordo1.5}, readily yields an upper bound for the approximation error, namely
\begin{equation}
\| (h,\mu_t^N)-(h,\mu_t) \|_p \le \sum_{k=0}^{T-1} \left\|
	\left(
		h, \Lambda_{t|t-k}(\mu_{t-k}^N)
	\right) - \left( 
		h, \Lambda_{t|t-k-1}(\mu_{t-k-1}^N)
	\right)
\right\|_p + \mS(h,T),
\label{eqGordo2}
\end{equation} 
and all we need to do is to calculate suitable bounds for the terms in the summation above.

It is not difficult to show (see Definition \ref{defLambda}) that, for any $\alpha \in \mP(D_\theta)$, 
\begin{equation}
\left(
	h, \Lambda_{t|t-k}(\alpha)
\right) = \frac{
	\left(
		h\prod_{j=0}^{k-1} u_{t-j}, \alpha
	\right)
}{
	\left(
		\prod_{j=0}^{k-1} u_{t-j}, \alpha
	\right)
},	
\label{eqGordo3}
\end{equation}
where $u_t(\theta) = (g_{t,\theta}^{y_t},\xi_{t,\theta})$. From \eqref{eqGordo3}, the $k$-th term in the summation of \eqref{eqGordo2} can be rewritten as 
\begin{equation}
\left(
	h, \Lambda_{t|t-k}(\mu_{t-k}^N)
\right) - \left( 
	h, \Lambda_{t|t-k-1}(\mu_{t-k-1}^N)
\right) = 
\frac{
	\left(
		h\prod_{j=0}^{k-1} u_{t-j}, \mu_{t-k}^N
	\right)
}{
	\left(
		\prod_{j=0}^{k-1} u_{t-j}, \mu_{t-k}^N
	\right)
}
- \frac{
\left(
		h\prod_{j=0}^{k-1} u_{t-j}, \Lambda_{t-k}(\mu_{t-k-1}^N)
	\right)
}{
	\left(
		\prod_{j=0}^{k-1} u_{t-j},\Lambda_{t-k}(\mu_{t-k-1}^N)
	\right)
} \nonumber
\end{equation}
hence, by way of inequality \eqref{eqPreliminaries}, we obtain
\begin{eqnarray}
\left\|
	\left(
		h, \Lambda_{t|t-k}(\mu_{t-k}^N)
	\right) - \left( 
		h, \Lambda_{t|t-k-1}(\mu_{t-k-1}^N)
	\right) 
\right\|_p &\le& \nonumber\\
a^k \left[
	\left\|
		\left(
			h\prod_{j=0}^{k-1} u_{t-j}, \mu_{t-k}^N
		\right) - \left(
			h\prod_{j=0}^{k-1} u_{t-j}, \Lambda_{t-k}( \mu_{t-k-1}^N )
		\right)
	\right\|_p 
\right. + && \nonumber\\
\left.
	\| h \|_\infty \left\|
		\left(
			\prod_{j=0}^{k-1} u_{t-j}, \mu_{t-k}^N
		\right) - \left(
			\prod_{j=0}^{k-1} u_{t-j}, \Lambda_{t-k}( \mu_{t-k-1}^N )
		\right)
	\right\|_p
\right], &&
\label{eqGordo4}
\end{eqnarray}
where we have made use of assumption A.\ref{asBounds_g-1} to obtain the factor $a^k$. 

The two $L_p$ norms on the right-hand side of \eqref{eqGordo4} have the form $\| (v,\mu_n^N) - (v,\Lambda_n(\mu_{n-1}^N) \|_p$, for $n=t-k$ and $v \in B(D_\theta)$ (namely, $v = h\prod_{j=0}^{k-1} u_{t-j}$ in the first term and $v = \prod_{j=0}^{k-1} u_{t-j}$ in the second term). Therefore, we now seek a bound for $\| (v,\mu_n^N) - (v,\Lambda_n(\mu_{n-1}^N) \|_p$ that can be substituted back into \eqref{eqGordo4}.

Recall that Algorithm \ref{alRA} succesively produces the approximate measures $\bar \mu_{n-1}^N = \frac{1}{N}\sum_{i=1}^N \delta_{\bar \theta_n^{(i)}}$, $\tilde \mu_n^N = \sum_{i=1}^N w_n^{(i)} \delta_{\bar \theta_n^{(i)}}$ and $\mu_n^N = \frac{1}{N}\sum_{i=1}^N \delta_{\theta_n^{(i)}}$. For the choice of kernel $\kappa_{N,\sfp}$ in \eqref{eqKappa} it is not difficult to show (see Appendix \ref{apMus}) that 
\begin{equation}
\| (v,\bar \mu_{n-1}^N) - (v, \mu_{n-1}^N) \|_p \le \frac{
	s_1 \| v \|_\infty
}{
	\sqrt{N}
},
\label{eqPasito1}
\end{equation}
where $s_1$ is a constant independent of $n$ and $N$, and
\begin{equation}
\left\|
	(v, \mu_n^N) - (v,\tilde \mu_n^N)
\right\|_p \le \frac{
	s_2\| v\|_\infty
}{
	\sqrt{N}
},
\label{eqPasito2}
\end{equation}
where $s_2$ is also constant w.r.t. $n$ and $N$ (note that $\mu_n^N$ is obtained from $\tilde \mu_n^N$ by way of a multinomial resampling step). Therefore, if we use the triangle inequality
\begin{eqnarray}
\| (v,\mu_n^N) - (v,\Lambda_n(\mu_{n-1}^N) \|_p &\le& 
\| (v,\mu_n^N) - (v,\tilde \mu_n^N) \|_p + \| (v,\tilde \mu_n^N) - (v,\Lambda_n(\bar \mu_{n-1}^N)) \|_p
\nonumber \\
&&+ \| (v,\Lambda_n(\bar \mu_{n-1}^N)) - (v,\Lambda_n(\mu_{n-1}^N)) \|_p
\label{eqTrianguloPasitos}
\end{eqnarray}
and realise that, by way of \eqref{eqPreliminaries} and assumption A.\ref{asBounds_g-1},
\begin{eqnarray}
\| (v,\Lambda_n(\bar \mu_{n-1}^N)) - (v,\Lambda_n(\mu_{n-1}^N)) \|_p &=&
\left\|
		\frac{
			(vu_n,\bar \mu_{n-1}^N)
		}{
			(u_n,\bar \mu_{n-1}^N)
		} - \frac{
			(vu_n,\mu_{n-1}^N)
		}{
			(u_n,\mu_{n-1}^N)
		}
\right\|_p \nonumber \\
&\le& a \| (vu_n,\bar \mu_{n-1}^N) - (u_n,\mu_{n-1}^N) \|_p 
+ a\| v \|_\infty \| (u_n,\bar \mu_{n-1}^N) - (u_n,\mu_{n-1}^N) \|_p, \nonumber\\
\label{eqTriangulito}
\end{eqnarray}
then it is straightforward to take \eqref{eqTriangulito}, \eqref{eqPasito1} and \eqref{eqPasito2} together and substitute them into \eqref{eqTrianguloPasitos} to obtain
\begin{equation}
\| (v,\mu_n^N) - (v,\Lambda_n(\mu_{n-1}^N) \|_p \le \frac{
	\|v \|_\infty( 2as_1 + s_2 )
}{
	\sqrt{N}
} + \| (v,\tilde \mu_n^N) - (v,\Lambda_n(\bar \mu_{n-1}^N)) \|_p
\label{eqTrianguloPasitos-2}
\end{equation}
and only the second term on the right hand side of the inequality above remains to be bounded. 

However, by the the construction of $\tilde \mu_n^N$ and Definition \ref{defLambda} (of $\Lambda_n$) we have
\begin{eqnarray}
\| (v,\tilde \mu_n^N) - (v,\Lambda_n(\bar \mu_{n-1}^N)) \|_p &=&
\left\|
	\frac{
		(vu_n^N, \bar \mu_{n-1}^N)
	}{
		(u_n^N, \bar \mu_{n-1}^N)
	} - \frac{
		(vu_n,\bar \mu_{n-1}^N)
	}{
		(u_n,\bar \mu_{n-1}^N)
	}
\right\|_p \nonumber \\
&\le& a \| (vu_n^N,\bar \mu_{n-1}^N) - (vu_n,\bar \mu_{n-1}^N) \|_p 
+ \| v \|_\infty a \| (u_N^N,\bar \mu_{n-1}^N) - (u_n,\bar \mu_{n-1}^N) \|_p. \nonumber\\
\label{eqCasiCasi}
\end{eqnarray}
Again, the two terms on the right hand side of the inequality \eqref{eqCasiCasi} have essentially the same form, hence it is enough to analyse the first one. Writing the integrals w.r.t. $\bar \mu_{n-1}^N$ explicitly, extracting $v \le \| v \|_\infty$ as a common factor and then applying Minkowski's inequality yields
\begin{equation}
\| (vu_n^N,\bar \mu_{n-1}^N) - (vu_n,\bar \mu_{n-1}^N) \|_p \le
\frac{
	\| v \|_\infty
}{
	N
} \sum_{i=1}^N  \| u_n^N(\bar \theta_n^{(i)}) - u_n(\bar \theta_n^{(i)} \|_p,
\nonumber 
\end{equation}
which, expanding the functions $u_n^N$ and $u_n$ as integrals w.r.t. $\xi_{n,\bar \theta_n^{(i)}}^N$ and $\xi_{n,\bar \theta_n^{(i)}}$, respectively, becomes
\begin{equation}
\| (vu_n^N,\bar \mu_{n-1}^N) - (vu_n,\bar \mu_{n-1}^N) \| \le
\frac{
	\| v \|_\infty
}{
	N
} \sum_{i=1}^N \left\| 
	(g_{n,\bar \theta_n^{(i)}}^{y_n}, \xi_{n,\bar \theta_n^{(i)}}^N) 
	- (g_{n,\bar \theta_n^{(i)}}^{y_n}, \xi_{n,\bar \theta_n^{(i)}}) 
\right\|_p.
\label{eqCasiCasi2}
\end{equation}
However, by assumption A.\ref{asBounds_g-1}, $\sup_{n \ge 0, \theta \in D_\theta, x \in \mX} g_{n,\theta}^{y_n} \le 1$, hence \eqref{eqCasiCasi2} can be extended as
\begin{equation}
\| (vu_n^N,\bar \mu_{n-1}^N) - (vu_n,\bar \mu_{n-1}^N) \|_p \le
\frac{
	\| v \|_\infty
}{
	N
} \sum_{i=1}^N \sup_{\ell \in B(\mX): \| \ell \|_\infty \le 1} \left(
	\sup_{n \ge 0}
		\left\| 
			(\ell, \xi_{n,\bar \theta_n^{(i)}}^N) - (\ell, \xi_{n,\bar \theta_n^{(i)}}) 
		\right\|_p,
\right)	
\label{eqCasiCasi3}
\end{equation}
where the terms $\sup_{n\ge 0} \| (\ell, \xi_{n,\bar \theta_n^{(i)}}^N) - (\ell, \xi_{n,\bar \theta_n^{(i)}}) \|_p$ can be controlled by way of Lemma \ref{lmUniformPhi}. To be specific, there exists a finite constant $\bar C$ 
independent of $N$ and $n$ such that 
\begin{equation}
\sup_{n \ge 0} \| (\ell, \xi_{n,\bar \theta_n^{(i)}}^N) - (\ell, \xi_{n,\bar \theta_n^{(i)}}) \|_p
\le \frac{
	\bar C
}{
	\sqrt{N}
}.
\label{eqMeollo}
\end{equation}
From \eqref{eqLaUltima} we readily see that there exists a constant $C^*<\infty$, independent of $n$, $N$, $a$ and $\ell$, such that $\bar C \le C^* \| \ell \|_\infty a^{3m+1}$, hence 
\begin{equation}
\sup_{\ell \in B(\mX): \|\ell\|_\infty \le 1} \bar C \le C^* a^{3m+1} < \infty.
\label{eqXurro0}
\end{equation}
Substituting \eqref{eqMeollo} back into \eqref{eqCasiCasi3} and using \eqref{eqXurro0} yields
\begin{equation}
\sup_{n \ge 0} \| (vu_n^N,\bar \mu_{n-1}^N) - (vu_n,\bar \mu_{n-1}^N) \|_p \le 
\frac{
	\| v \|_\infty C^* a^{3m+1}
}{
	\sqrt{N}
}.
\label{eqArrecu1}
\end{equation}

From \eqref{eqArrecu1}, we can substitute back into the sequence of inequalities that starts at \eqref{eqGordo2}. In particular, inserting \eqref{eqArrecu1} into \eqref{eqCasiCasi} yields
\begin{equation}
\sup_{n\ge 0} \| (v,\tilde \mu_n^N) - (v,\Lambda_n(\bar \mu_{n-1}^N)) \|_p
\le \frac{
	2\|v\|_\infty C^* a^{3m+2}
}{
	\sqrt{N}
}
\label{eqArrecu2}
\end{equation}
and plugging \eqref{eqArrecu2} into \eqref{eqTrianguloPasitos-2} we arrive at 
\begin{equation}
\sup_{n\ge 0} \| (v,\mu_n^N) - (v,\Lambda_n(\mu_{n-1}^N) \|_p  \le 
\frac{
	\| v \|_\infty \tilde C^* a^{3m+2}
}{
	\sqrt{N}
},
\label{eqArrecu3}
\end{equation}
where $\tilde C^* = 2C^*+2s_1+s_2$. The expression above yields bounds for the two terms on the right hand side of \eqref{eqGordo4}. Hence, substituting \eqref{eqArrecu3} into \eqref{eqGordo4} we can write
\begin{eqnarray}
\left\|
	\left(
		h, \Lambda_{t|t-k}(\mu_{t-k}^N)
	\right) - \left( 
		h, \Lambda_{t|t-k-1}(\mu_{t-k-1}^N)
	\right) 
\right\|_p 
&\le& \frac{
	2 \| h \|_\infty \tilde C^* a^{3m+2+k}
}{
	\sqrt{N}
}
\label{eqArrecu6}
\end{eqnarray}
The inequality \eqref{eqArrecu6}, in turn, provides bounds for each one of the terms in the summation of \eqref{eqGordo2} which, taken together, lead to 
\begin{eqnarray}
\sup_{t \ge T} \| (h,\mu_t^N)-(h,\mu_t) \|_p &\le&
\frac{
	\| h \|_\infty \hat C T a^T
}{
	\sqrt{N}
} + \mS(h,T),
\label{eqArrecu7}
\end{eqnarray}
where $\hat C = 2 \tilde C^* a^{3m+2}$. 

Next, we prove that a bound of the form in \eqref{eqArrecu7} also holds for $t < T$. In this case we can decompose the $L_p$ norm of the approximation error as
\begin{eqnarray}
\| (h,\mu_t^N) - (h,\mu_t) \|_p &\le& 
\sum_{k=0}^{t-1} \| (h, \Lambda_{t|t-k}(\mu_{t-k}^N)) - (h, \Lambda_{t|t-k-1}(\mu_{t-k-1}^N)) \|_p
+ \| (h, \Lambda_{t|0}(\mu_0^N) - (h,\Lambda_{t|0}(\mu_0)) \|_p.
\nonumber \\
\label{eqCachumbre}
\end{eqnarray}
The sum on the right hand side of \eqref{eqCachumbre} has the same structure as the summation in  \eqref{eqGordo2}, hence exactly the same argument leading to \eqref{eqArrecu7} (and bearing in mind that $t<T$) yields
\begin{eqnarray}
\sup_{t < T} \sum_{k=0}^{t-1} \| (h, \Lambda_{t|t-k}(\mu_{t-k}^N)) - (h, \Lambda_{t|t-k-1}(\mu_{t-k-1}^N)) \|_p
&\le& \frac{
	\| h \|_\infty \hat C T a^T
}{
	\sqrt{N}
}
\label{eqCachu1}
\end{eqnarray}
which is the same bound as in \eqref{eqArrecu7} except for the residual $\mS(h,T)$. As for the last term in \eqref{eqCachumbre}, recall from \eqref{eqGordo3} that $(h,\Lambda_{t|0}(\alpha)) = ( h\prod_{j=0}^{t-1} u_{t-j}, \alpha) /  ( \prod_{j=0}^{t-1} u_{t-j}, \alpha)$ which, combined with \eqref{eqPreliminaries}, yields
\begin{eqnarray}
\left\|
	(h, \Lambda_{t|0}(\mu_0^N)) - (h,\Lambda_{t|0}(\mu_0)) 
\right\|_p &\le& a^t \left[
	\left\|
		\left(
			h\prod_{j=0}^{t-1} u_{t-j}, \mu_0^N
		\right) - \left(
			h\prod_{j=0}^{t-1} u_{t-j}, \mu_0 )
		\right)
	\right\|_p 
\right. \nonumber\\
&& + \left.
	\| h \|_\infty \left\|
		\left(
			\prod_{j=0}^{t-1} u_{t-j}, \mu_0^N
		\right) - \left(
			\prod_{j=0}^{t-1} u_{t-j}, \mu_0^N )
		\right)
	\right\|_p
\right].
\nonumber
\end{eqnarray}
Since $\mu_0^N$ is a random measure constructed with $N$ i.i.d. samples from the distribution with measure $\mu_0$, it is straightforward to show that there is a constant $\bar c_0<\infty$, independent of $N$ and $t$, such that
\begin{equation}
\left\|
	(h, \Lambda_{t|0}(\mu_0^N)) - (h,\Lambda_{t|0}(\mu_0)) 
\right\|_p \le \frac{
	a^t \| h \|_\infty \bar c_0
}{
	\sqrt{N}
}.
\label{eqCu1}
\end{equation}
If we recall that $t < T$ and put together \eqref{eqCachumbre}, \eqref{eqCachu1} and \eqref{eqCu1} then we readily obtain the bound
\begin{equation}
\sup_{t < T} \| (h,\mu_t^N) - (h,\mu_t) \|_p \le \frac{
	\sfC \| h \|_\infty  T a^T
}{
	\sqrt{N}
}, \label{eqCu2}
\end{equation}
where $\sfC = \hat C \vee \bar c_0$ is a finite constant independent of $N$, $T$ and $h$. 

Combining the inequalities \eqref{eqArrecu7} and \eqref{eqCu2} we have the error bound
\begin{equation}
\sup_{t \ge 0} \| (h,\mu_t^N) - (h,\mu_t) \|_p \le \frac{
	\sfC \| h \|_\infty  T a^T
}{
	\sqrt{N}
} + \mS(h,T)
\label{eqCu3}
\end{equation}
that holds for any positive integer $T<\infty$. In particular, we can choose $T=T_N^\epsilon$ such that $\sfC \| h \|_\infty T_N^\epsilon a^{T_N^\epsilon} \le N^\epsilon$ for any $0 < \epsilon < \frac{1}{2}$. It is sufficient to set
\begin{equation}
T_N^\epsilon = \left\lfloor 
	\frac{
		\epsilon \log(N) - \log(\sfC\|h\|_\infty)
	}{
		1 + \log(a)
	}
\right\rfloor
\label{eqChoiceOfT}
\end{equation}
in order to substitute $T=T_N^\epsilon$ in \eqref{eqCu3} and obtain
\begin{equation}
\sup_{t \ge 0} \| (h,\mu_t^N) - (h,\mu_t) \|_p \le \frac{
	1
}{
	N^{\frac{1}{2}-\epsilon}
} + \mS(h,T_N^\epsilon).
\label{eqCu4}
\end{equation}
Since $\lim_{N\rw\infty} T_N^\epsilon = \infty$ for every $\epsilon \in (0,\frac{1}{2})$, then assumption A.\ref{asStabilityLambda} implies that 
$$
\lim_{N\rw\infty} \mS(h,T_N^\epsilon) = 0
$$ 
and, as a consequence of \eqref{eqCu4}, $\lim_{N\rw\infty}  \sup_{t \ge 0} \| (h,\mu_t^N) - (h,\mu_t) \|_p = 0$.

To complete the proof, we observe that assumption A.\ref{asStabilityLambda-b} combined with \eqref{eqChoiceOfT} yields
\begin{equation}
\mS(h,T_N^\epsilon) \le C N^{-\epsilon \frac{
		\bar b_2
	}{
		1 + \log(a)
	}
},
\label{eqPoorRate3}
\end{equation}
where $C=( \sfC \|h\|_\infty )^\frac{\bar b_2 }{1+\log(a)}<\infty$ is independent of $N$ and $t$. Combining \eqref{eqPoorRate3} with \eqref{eqCu4} yields the explicit error bound in the statement of Theorem \ref{thUniform}.

$\QED$

\begin{Nota}
While the convergence of Algorithm \ref{alRA} can be guaranteed without assumption A.\ref{asStabilityLambda-b}, the latter is necessary in order to obtain the error bound in the statement of Theorem \ref{thUniform}. To be specific, we need to specify how fast the error $\mS(h,T)$ vanishes in order to compute an explicit error bound. This is given by assumption A.\ref{asStabilityLambda-b}, which describes a feature of the state-space model (rather than a feature of the algorithm).
\end{Nota}
\subsection{Parameter identification} \label{ssIdentification}

The uniform convergence result of Theorem \ref{thUniform} implies that the vector of model parameters can be estimated exactly (as $t \rw\infty$) provided that the sequence of observations is informative enough to guarantee that the posterior probability mass asymptotically concentrates around a single point in the parameter space $D_\theta$. To be specific, in this section we assume that there exists $\theta_* \in D_\theta$ (which may be thought of as the ``true value´´ of $\Theta$) such that
\begin{equation}
\lim_{t\rw\infty} \mu_t = \delta_{\theta_*}
\label{eqIdentif1}
\end{equation}
for the available sequence of observations $\{ y_t \}_{t > 0}$ and then proceed to show that $\mu_t^N \rw \delta_{\theta_*}$ as $t\rw\infty$, in a sense to be made precise. The existence of such $\theta_*$ is not a strong assumption. In \cite{Papavasiliou06} it is shown that, provided the parameter is ``identifiable'', meaning that
\begin{equation}
\theta_1 = \theta_2 \Leftrightarrow \lim_{t\rw\infty} \phi_{t,\theta_1} = \lim_{t\rw\infty} \phi_{t,\theta_2},
\nonumber 
\end{equation}
then the limit in \eqref{eqIdentif1} holds a.s. under mild assumptions. 

Let $\Omega = \{ h_i \in B(D_\theta): \| h_i \|_\infty \le 1, i \ge 1 \}$ be a convergence determining set \cite[Theorem 2.18]{Bain08} and define the distance $d_\Omega:\mP(D_\theta) \times \mP(D_\theta) \rw [0,+\infty)$ as
\begin{equation}
d_\Omega(\alpha,\eta) \dfn \sum_{i \ge 1} \frac{1}{2^i} |(h_i,\alpha) - (h_i,\eta)|
\nonumber
\end{equation}
for any $\alpha, \eta \in \mP(D_\theta)$. The existence of $\Omega$ is granted by \cite[Theorem 2.18]{Bain08}, while  \cite[Theorem 2.19]{Bain08} shows that a sequence of measures $\{ \alpha_t \in \mP(D_\theta) \}_{t \ge 1}$, converges weakly to another measure $\alpha \in \mP(D_\theta)$ if, and only if,
$
\lim_{t\rw\infty} d_\Omega(\alpha_t,\alpha) = 0 
$
. The following result regarding the asymptotic identification of the system parameters is a fairly direct consequence of Theorem \ref{thUniform}.

\begin{Teorema} \label{thIdentification}
Let $D_\theta$ be a compact set and $\kappa_{N,\sfp}$ a kernel of the class in Eq. \eqref{eqKappa}. If assumptions A.\ref{asStabilityLambda}--A.\ref{asPsiLipschitz} hold and there exists $\theta_*\in D_\theta$ such that $\lim_{t\rw\infty} \mu_t=\delta_{\theta_*}$, then, for any $0 < \epsilon < \frac{1}{2}$,
\begin{equation}
\limsup_{t\rw\infty} E\left[
	d_\Omega(\mu_t^N,\delta_{\theta_*})
\right] \le N^{-\frac{1}{2}+\epsilon} + C N^{-\epsilon\frac{\bar b_2}{1 + \log(a)}} + 2^{-N+1}
\label{eqIdentif1.3}
\end{equation}
where $C$, $\bar b_2$ and $a$ are finite constants independent of $N$ and $t$. In particular, 
$$
\lim_{N\rw\infty} \limsup_{t\rw\infty} E\left[
	d_\Omega(\mu_t^N,\delta_{\theta_*})
\right] = 0.
$$
\end{Teorema}

\noindent \textit{\textbf{Proof:}} We start with the triangle inequality
\begin{equation}
\sup_{n\ge t} E\left[
	d_\Omega(\mu_n^N,\delta_{\theta_*})
\right] \le \sup_{n \ge t} \left(
	E\left[
		d_\Omega(\mu_n^N,\mu_n) 
	\right] + d_\Omega(\mu_n,\delta_{\theta_*})
\right).
\label{eqIdentif1.5}
\end{equation}
If we choose an integer $K \ge 1$ and expand $d_\Omega$, the term $d_\Omega(\mu_n^N,\mu_n)$ can be upper bounded as
\begin{eqnarray}
d_\Omega(\mu_n^N,\mu_n) &=& 
\sum_{i=1}^K \frac{1}{2^i} | (h_i,\mu_n^N) - (h_i,\mu_n) | + \sum_{j>K} \frac{1}{2^j} | (h_i,\mu_n^N) - (h_i,\mu_n) | \nonumber\\
&\le& \sum_{i=1}^K \frac{1}{2^i} | (h_i,\mu_n^N) - (h_i,\mu_n) | + \frac{1}{2^{K-1}},
\label{eqIdentif2}
\end{eqnarray}
where the inequality follows from bounding $| (h_i,\mu_n^N) - (h_i,\mu_n) | \le 2$ and then computing $\sum_{j>K} 2^{-j}= 2^{-K}$. From \eqref{eqIdentif2}, we readily obtain
\begin{eqnarray}
\sup_{n\ge t} E\left[
	d_\Omega(\mu_n^N,\mu_n) 
\right] &\le& \sum_{i=1}^K \frac{1}{2^i} \sup_{n \ge t} \| (h_i,\mu_n^N)-(h_i,\mu_n) \|_1 + \frac{1}{2^{K-1}}
\nonumber \\
&\le& e(N) \left( 1 - \frac{1}{2^{K-1}} \right) + \frac{1}{2^{K-1}},
\label{eqLastOne}
\end{eqnarray}
where we have applied the identity $\sum_{i=1}^K 2^{-i} = 1 - 2^{-K+1}$ and the inequality
\begin{equation}
\sup_{n \ge t} \| (h_i,\mu_n^N)-(h_i,\mu_n) \|_1 \le N^{-\frac{1}{2}+\epsilon} + C N^{-\epsilon\frac{\bar b_2}{1 + \log(a)}} \dfn e(N).
\label{eqIdentif3}
\end{equation}
The latter follows from Theorem \ref{thUniform}, with arbitrary $\epsilon \in (0,\frac{1}{2})$ and finite constants $C$, $\bar b_2$ and $a$ independent of $N$ and $t$. The inequality \eqref{eqLastOne} is valid for any $K$, hence if we choose $N=K$ it readily follows that 
\begin{equation}
\sup_{n \ge t} E\left[
	d_\Omega( \mu_n^N, \mu_n )
\right] \le e(N) + 2^{-N+1}.
\label{eqIdentif4}
\end{equation}

If we now substitute \eqref{eqIdentif4} into \eqref{eqIdentif1.5} we obtain
\begin{equation}
\sup_{n \ge t} E\left[
	d_\Omega(\mu_n^N,\delta_{\theta_*})
\right] \le e(N) + 2^{-N+1} + \sup_{n \ge t} d_\Omega(\mu_n, \delta_{\theta_*})
\nonumber
\end{equation}
and taking the limit as $t\rw\infty$ yields
\begin{equation}
\limsup_{t\rw\infty} E\left[
	d_\Omega(\mu_t^N,\delta_{\theta_*})
\right] \le e(N) + 2^{-N+1},
\nonumber
\end{equation}
since $\lim_{t\rw\infty} \mu_t=\delta_{\theta_*}$ by assumption. Finally, note that $e(N)+2^{-N+1}$ is exactly the bound in \eqref{eqIdentif1.3}.
 
$\QED$
\section{Numerical results} \label{sExamples}

\subsection{Simulation setup}

We present some computer simulation results to illustrate the numerical performance of the proposed nested particle filtering scheme (Algorithm \ref{alRA}) with long sequences of observations. A study numerical of convergence with increasing number of particles is presented in \cite{Crisan15}. Let us consider a 3-dimensional Lorenz system \cite{Lorenz63} with additive dynamical noise and partial noisy observations \cite{Chorin04}. The state of this system is a 3-dimensional stochastic process $\{ X(s) \}_{s\in(0,\infty)}$, taking values on $\Real^3$, which evolves over time according to the stochastic differential equations
$$
dX_1 = -S(X_1-Y_1)ds + dW_1, \quad 
dX_2 = \left( RX_1 - X_2 - X_1X_3 \right)ds + dW_2, \quad 
dX_3 = \left( X_1X_2 - BX_3 \right)ds + dW_3, \nonumber
$$
where $\{ W_i(s) \}_{s\in(0,\infty)}$, $i=1, 2, 3$, are independent 1-dimensional Wiener processes and $(S,R,B)\in\Real$ are unknown model parameters. To put this system within the framework of this paper, we apply Euler's method with integration step $\Delta>0$ to obtain the stochastic difference equations
\begin{eqnarray}
X_{1,t} &=& X_{1,t-1} - \Delta S(X_{1,t-1}-X_{2,t-1}) + \sqrt{\Delta} U_{1,t},\label{eqDiscreteLorenz-1}\\
X_{2,t} &=& X_{2,t-1} + \Delta ( RX_{1,t-1} - X_{2,t-1} - X_{1,t-1}X_{3,t-1} ) + \sqrt{\Delta} U_{2,t}, \label{eqDiscreteLorenz-2}\\
X_{3,t} &=& X_{3,t-1} + \Delta ( X_{1,t-1}X_{2,t-1} - BX_{3,t-1} ) + \sqrt{\Delta} U_{3,t}, \label{eqDiscreteLorenz-3}
\end{eqnarray} 
where $\{ U_{i,t} \}_{t=0, 1, ...}$, $i=1,2,3$, are independent sequences of i.i.d. normal r.v.'s with 0 mean and variance 1. The system is partially observed every 40 discrete-time steps, and the observations have the form  $\{ Y_n=(Y_{1,n},Y_{3,n}) \}_{n=1, 2, ...}$, where
\begin{equation}
Y_{1,n} = k_o X_{1,40n} + V_{1,n}, \quad
Y_{3,n} = k_o X_{3,40n} + V_{3,n}, \label{eqObservLorenz}
\end{equation}
$k_o >0$ is an unknown scale parameter and $\{ V_{i,n} \}_{n=1, 2, ...}$, $i=1,3$, are independent sequences of i.i.d. normal random variables with zero mean and variance $\sigma^2 = \frac{1}{10}$.

Let $X_t=(X_{1,t},X_{2,t},X_{3,t})$ be the state vector, let $Y_n=(Y_{1,n},Y_{3,n})$ be the observation vector and let $\Theta = (S,R,B,k_o)$ be the static and unknown model parameters to be estimated. It is simple to obtain the family of kernels $\tau_{t,\theta}(dx|x_{t-1})$ from Eqs. \eqref{eqDiscreteLorenz-1}--\eqref{eqDiscreteLorenz-3} and the likelihood $g_{n,\theta}^{y_n}(x_n)$ from Eq. \eqref{eqObservLorenz}.  The sequences $X_t$ and $Y_n$ are defined on different time scales, however it is straightforward to construct a sequence $\hat X_n$, with the same time index as the observations, if we simply define $\hat X_n = X_{40n}$. The transition kernel for $\hat X_n$ is obtained by composing the kernels for $X_t$. In particular, for the purpose of implementing Algorithm \ref{alRA}, one can draw a sample $\hat X_n = \hat x_n$ conditional on $\theta$ and $\hat X_{n-1} = \hat x_{n-1}$, by successively simulating
$$
\tilde x_t \sim \tau_{t,\theta}(dx|\tilde x_{t-1}), \quad t=40(n-1)+1, ..., 40n, 
$$
where $\tilde x_{40(n-1)} = \hat x_{n-1}$ and $\hat x_n = \tilde x_{40n}$. The prior measure for the state variables is normal and independent of $\Theta$, namely
$$
X_0 \sim \mN(x_*,v_0^2 \mI_3),
$$
where $x_* = (-5.91652; -5.52332; 24.5723)$ is the mean and $v_0^2\mI_3$ is the covariance matrix, with $v_0^2 = 10$ and $\mI_3$ the 3-dimensional identity matrix. The value $x_*$ has been taken from a simulated trajectory of the deterministic Lorenz 63 model. In this way we ensure that the simulation for the stochastic model starts at a ``reasonable'' point in the state space.

The goal is to track the posterior probability measures of the parameters, $\mu_n(d\theta) = \Prob\{ \Theta \in d\theta | Y_{1:n} \}$, $n=1, 2, ...$, using Algorithm \ref{alRA}. We assume that the parameters are a priori independent, namely
$$
S \sim \mU(5, 20), \quad R \sim \mU(18, 50), \quad B \sim \mU(1,8) \quad \mbox{and} \quad k_o \in \mU(0.5, 3),
$$ 
where $\mU(a,b)$ is the uniform probability distribution in the interval $(a,b)$. Therefore the prior measure $\mu_0$ is uniform, with support $D_\theta = [5,20] \times [18,50] \times [18,50] \times [1,8] \times [0.5, 3]$. 

In order to run Algorithm \ref{alRA} we need to choose the number of particles in the state space, $N$, the number of particles in the parameter space $M$, and the jittering kernel $\kappa_{N,\sfp}$. For the set of computer experiments here, we have set $N=M=300$ and the jittering kernel is selected as in \eqref{eqKappa}, in particular
$$
\kappa_{N,\sfp}^{\theta_{n-1}}(d\theta) = (1-\epsilon_N)\delta_{\theta_{n-1}}(d\theta) + \epsilon_N \bar \kappa^{\theta_{n-1}}(\theta)d\theta,
$$
where $\epsilon_N = \frac{1}{\sqrt{N}}$ and $\bar \kappa^{\theta_{n-1}}(\theta)$ is a truncated-Gaussian pdf with support $D_\theta$ and independent of $N$, namely
$$
\bar \kappa^{\theta_{n-1}}(\theta) = {\sf c}_{n-1} \exp\left\{
	-\frac{1}{2}\left(\theta - \theta_{n-1} \right)^\top 
	\mC^{-1} 
	\left(\theta - \theta_{n-1} \right)
\right\}, \quad \theta \in D_\theta,
$$
where the proportionality constant ${\sf c}_{n-1}$ is a function of $\theta_{n-1}$ and the (fixed) covariance matrix is 
$$
\mC = \left[
	\begin{array}{cccc}
	\frac{1}{2} &0 &0 &0\\
	0 &\frac{1}{2} &0 &0\\
	0 &0 &\frac{1}{5} &0\\
	0 &0 &0 &\frac{1}{20}\\
	\end{array}
\right].
$$

\subsection{Results}

The actual parameter values used for the computer experiments in this section are $(S,R,B,k_o) = (10,28,\frac{8}{3},0.8)$, which yield an underlying chaotic dynamics.

\begin{figure}
\centerline{
        \subfigure[Posterior-mean estimates of parameter $S$.]{
                \includegraphics[width=0.35\linewidth]{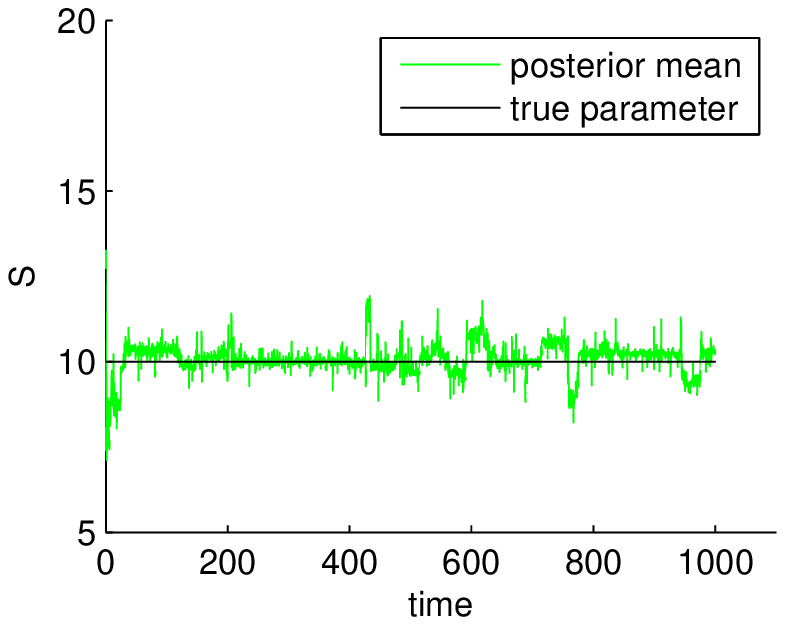}
        }
        \subfigure[Posterior-mean estimates of parameter $R$.]{
                \includegraphics[width=0.35\linewidth]{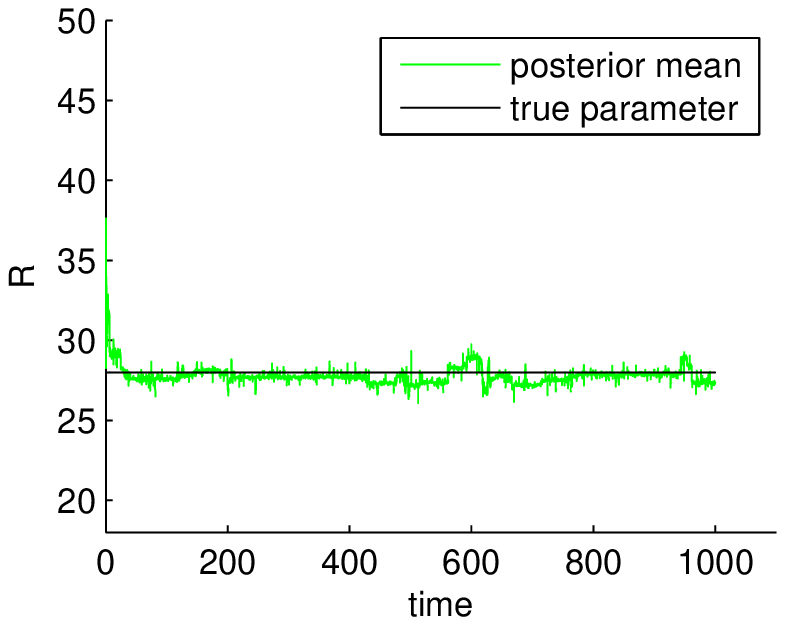}
        }
}
\centerline{
        \subfigure[Posterior-mean estimates of parameter $B$.]{
                \includegraphics[width=0.35\linewidth]{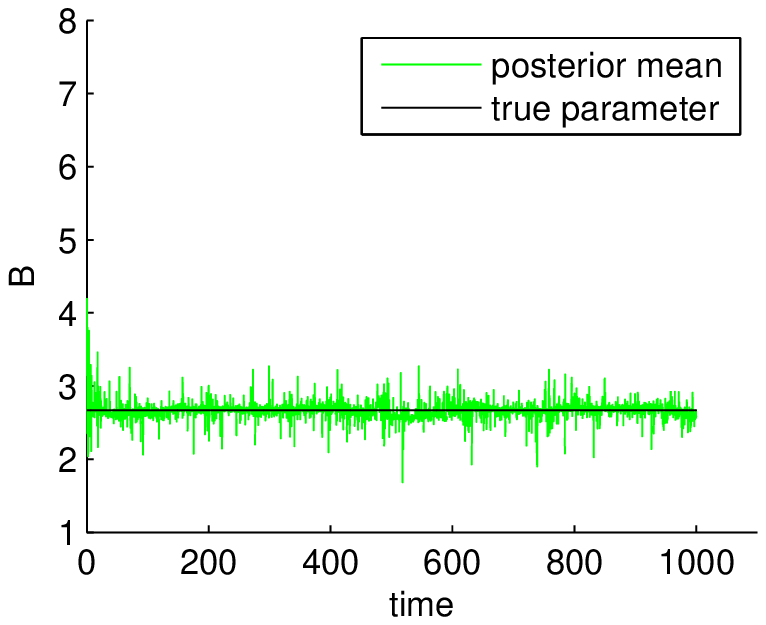}
        }
        \subfigure[Posterior-mean estimates of parameter $k_o$.]{
                \includegraphics[width=0.35\linewidth]{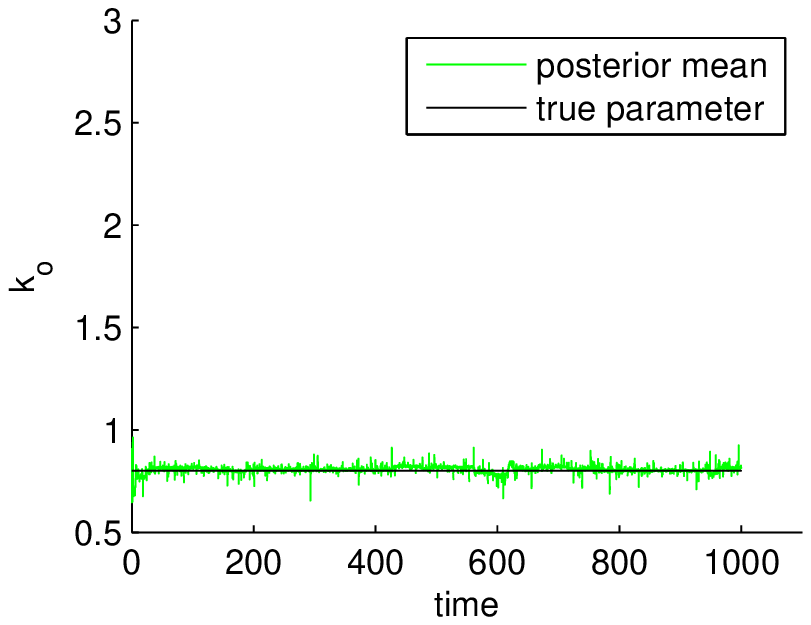}
        }
}
\caption{Evolution of the posterior-mean estimates of the Lorenz 63 model parameters $S,R,B$ and $k_o$ over time. The horizontal axes are labeled with continuous time units. After Euler's discretisation, each continuous time unit amounts to 1,000 discrete time steps (hence, 1 million time steps for the complete simulation), with one observation vector every 40 discrete-time steps. The number of particles is $N=M=300$. The vertical axes extend over the exact prior support for each parameter, i.e., $S \in [5, 20], R\in[18,50], B\in[1,8]$ and $k_o\in[0.5,3]$.}
\label{fRealisation_SRBk_t1000}
\end{figure}

Figure \ref{fRealisation_SRBk_t1000} shows the posterior mean estimates of the parameters $S,R,B$ and $k_o$ obtained for a single simulation with $N=M=300$ particles and a length of 1,000 continuous time units. Since the Euler's integration step is $\Delta=10^{-3}$ continuous time units and observations are taken every $40\Delta$ continuous time units, the simulation involves $10^6$ discrete time steps and $25\times 10^3$ observations vectors. At discrete time $n$, the posterior mean of the parameter vector $\Theta=(S,R,B,k_o)$ is computed as $\hat \theta_n^N = \frac{1}{N}\sum_{i=1}^N w_n^{(i)} \bar \theta_n^{(i)}$. In the same figure it can be seen that, after a relatively short convergence period, the estimates remain locked to the true parameter values (plotted with black solid lines). The posterior-mean approximation $\hat \theta_n^N$ is random and it only converges to the exact posterior mean as $N\rw\infty$, hence some fluctuations can be observed over time. However, the amplitude of the fluctuations remains bounded and stable over the whole simulation run. 
  
\begin{figure}
\centerline{
        \subfigure[Normalised posterior standard deviation of the parameter $S$.]{
                \includegraphics[width=0.35\linewidth]{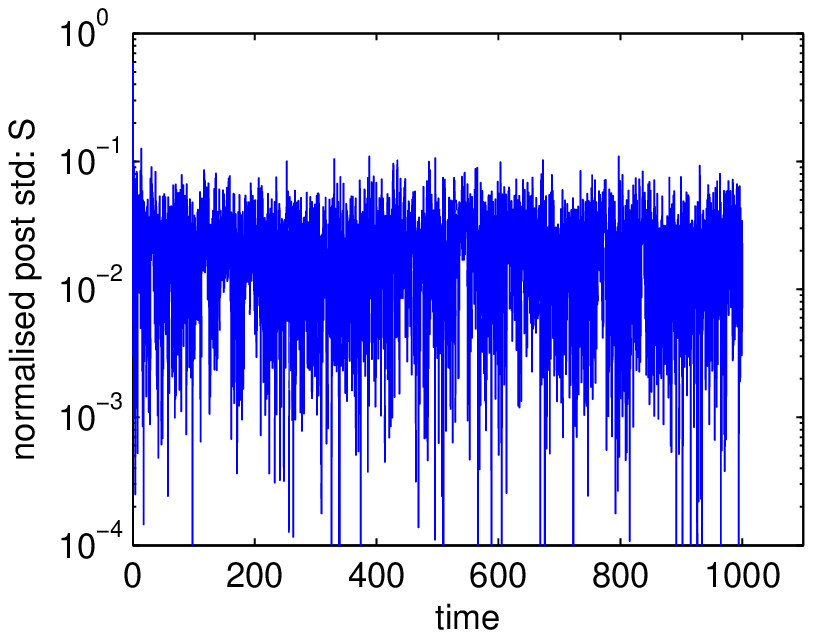}
        }
        \subfigure[Normalised posterior standard deviation of the parameter $R$.]{
                \includegraphics[width=0.35\linewidth]{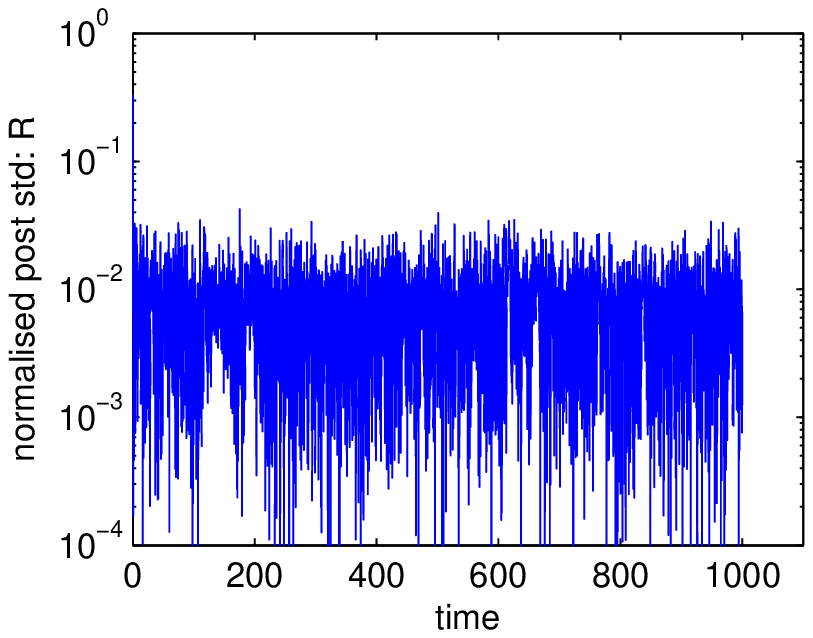}
        }
}
\centerline{
        \subfigure[Normalised posterior standard deviation of the parameter $B$.]{
                \includegraphics[width=0.35\linewidth]{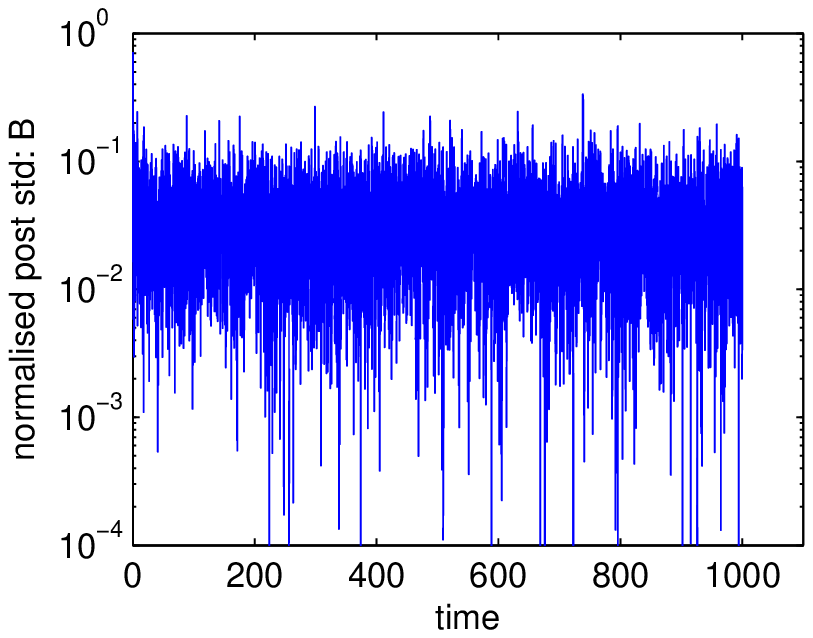}
        }
        \subfigure[Normalised posterior standard deviation of the parameter $k_o$.]{
                \includegraphics[width=0.35\linewidth]{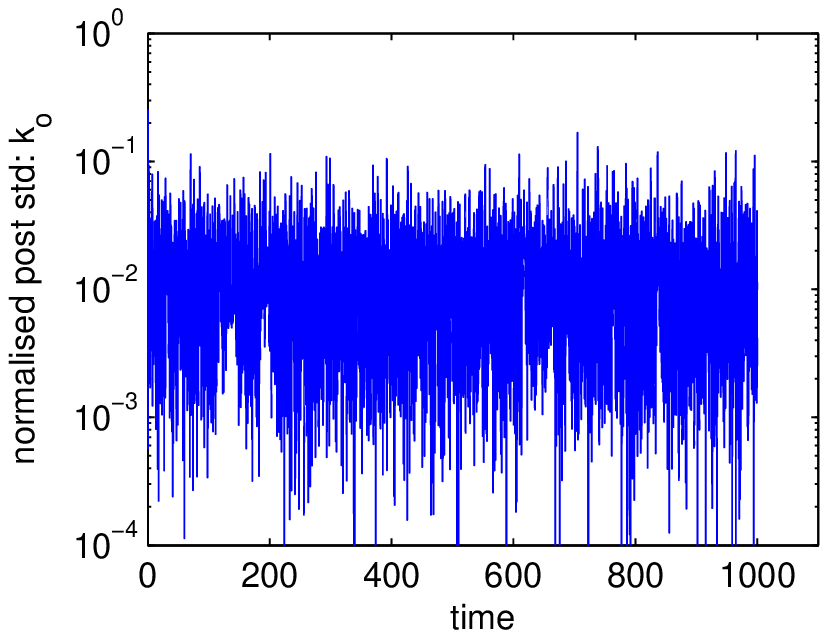}
        }
}
\caption{Evolution of the normalised posterior standard deviation of the Lorenz 63 model parameters $S,R,B$ and $k_o$ over time. The horizontal axes are labeled with continuous time units. After Euler's discretisation, each continuous time unit amounts to 1,000 discrete time steps, with one observation vector every 40 discrete-time steps. The number of particles is $N=M=300$.}
\label{fSTD_t1000}
\end{figure}

Figure \ref{fSTD_t1000} shows the normalised posterior standard deviation (NSTD) of the parameter estimates for the same simulation run. At each time $n$, this is computed for the $j$-th parameter, $j=1, ..., 4$, as 
$$
NSTD_{j,n} = \frac{
	\sqrt{ 
		\sum_{i=1}^N w_n^{(i)} (\bar \theta_{j,n}^{(i)} - \hat \theta_{j,n}^N)^2 
	}
}{
	\theta_j^*
},
$$
where $\theta_j^*$ is the true value of the $j$-th parameter (namely, $\theta_1^*=S=10, \theta_2^*=R=28, \theta_3^*=B=\frac{8}{3}$ and $\theta_4^*=k_o=0.8$). Again, the NSTD is a random statistic and it displays fluctuations, however it can be seen that their amplitudes remain bounded and there is no apparent increase over time.

\begin{figure}
\centerline{
        \subfigure[Error $e_{1,n}^N = \hat x_{1,n}^N - \hat x_{1,n}$.]{
                \includegraphics[width=0.35\linewidth]{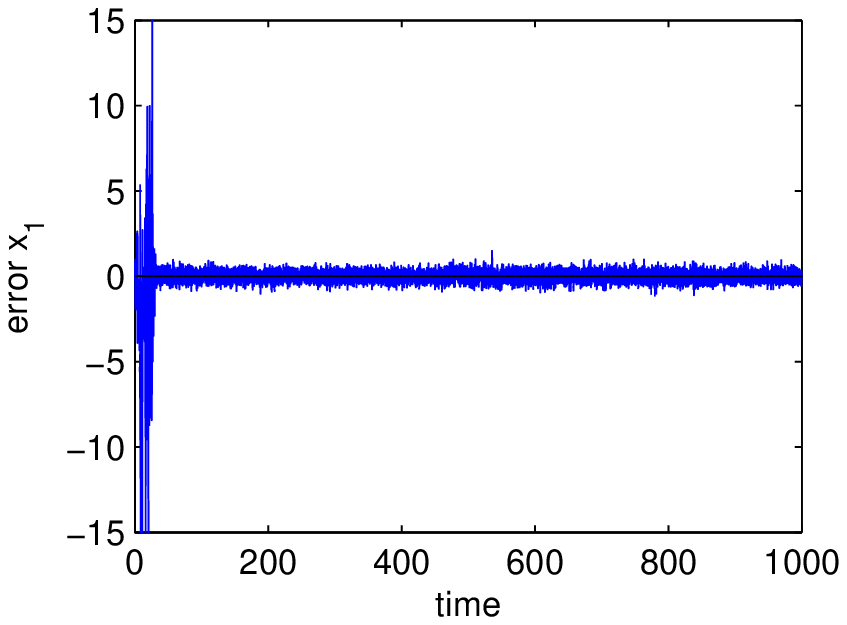}
        }
        \subfigure[Error $e_{2,n}^N = \hat x_{2,n}^N - \hat x_{2,n}$.]{
                \includegraphics[width=0.35\linewidth]{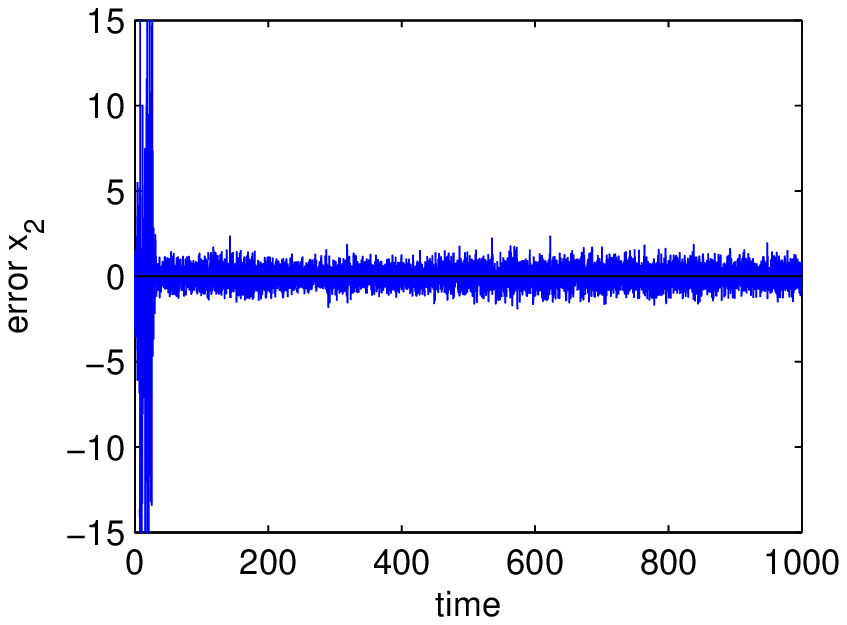}
        }
}
\centerline{
        \subfigure[Error $e_{3,n}^N = \hat x_{3,n}^N - \hat x_{3,n}$.]{
                \includegraphics[width=0.35\linewidth]{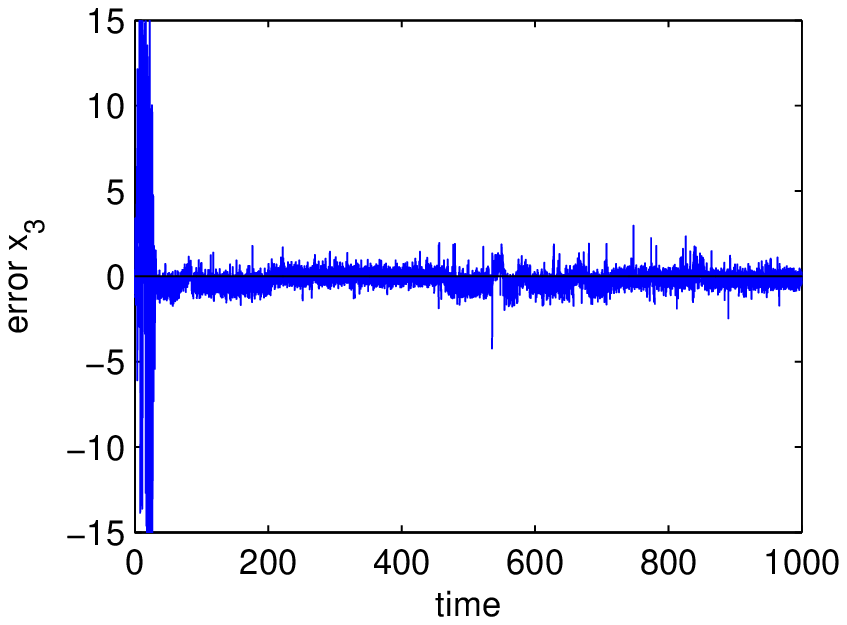}
        }
}
\caption{Evolution of the errors $e_{\ell,n}^N = \hat x_{\ell,n}^N - \hat x_n$, $\ell=1,2,3$, for the state variables of the Lorenz 63 model, where the estimates $\hat x_{\ell,n}^N$ are posterior means. The horizontal axes are labeled with continuous time units. After Euler's discretisation, each continuous time unit amounts to 1,000 discrete time steps (hence, 1 million time steps for the complete simulation), with one observation vector every 40 discrete-time steps. The number of particles is $N=M=300$.}
\label{fX_t1000}
\end{figure}

Figure \ref{fX_t1000} displays the errors between the posterior-mean estimates of the state variables and the actual values, for the same simulation run as in Figures \ref{fRealisation_SRBk_t1000} and \ref{fSTD_t1000}. At discrete time $n$, the estimates are computed as $\hat x_{\ell,n}^N = \frac{1}{N} \sum_{i=1}^N  w_n^{(i)} \sum_{j=1}^N \hat x_n^{(i,j)}$, for $\ell = 1, 2, 3$, and the errors displayed are of the form $e_{\ell,n}^N = \hat x_{\ell,n}^N - \hat x_n$. It can be seen that the errors are large at the beginning of the simulation. This is a consequence of the initial uncertainty in the values of the fixed parameters. Once the parameter estimates have converged, the errors decrease substatially and remain bounded, stable and centred around 0 for the rest of the simulation. 

Finally, we have carried out a set of 50 independent simulations in order to approximate the mean absolute error of the parameter (posterior-mean) estimates. For each simulation we have run the stochastic Lorenz 63 model for 400 continuous time units, which amounts to $400 \times 10^3$ discrete time steps and a sequence of $10,000$ observations. For each simulation and each time step, we have computed the absolute error of the posterior-mean estimate of each parameter. Then, we have averaged these errors over the 50 independent simulation runs.

\begin{figure}
\centerline{
        \subfigure[Average absolute error of the posterior-mean estimates; parameter $S$.]{
                \includegraphics[width=0.35\linewidth]{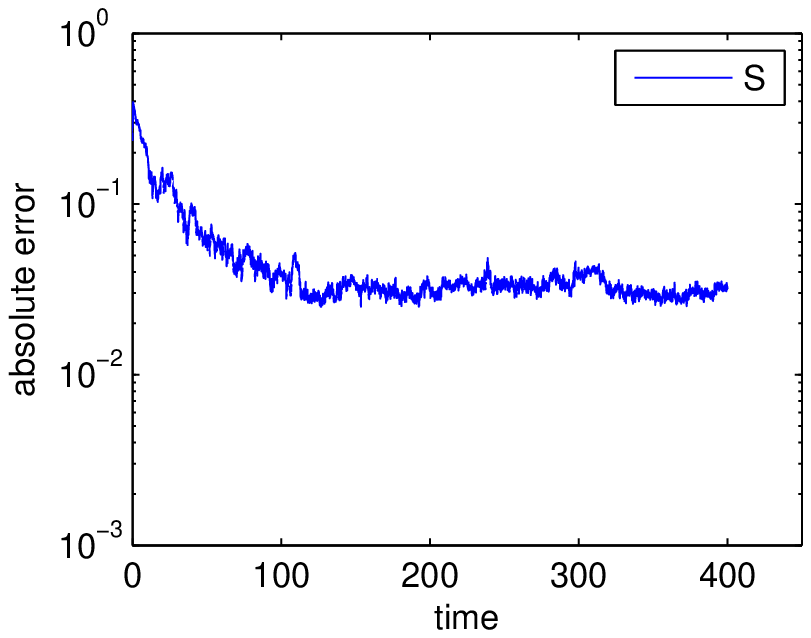}
        }
        \subfigure[Average absolute error of the posterior-mean estimates; parameter $R$.]{
                \includegraphics[width=0.35\linewidth]{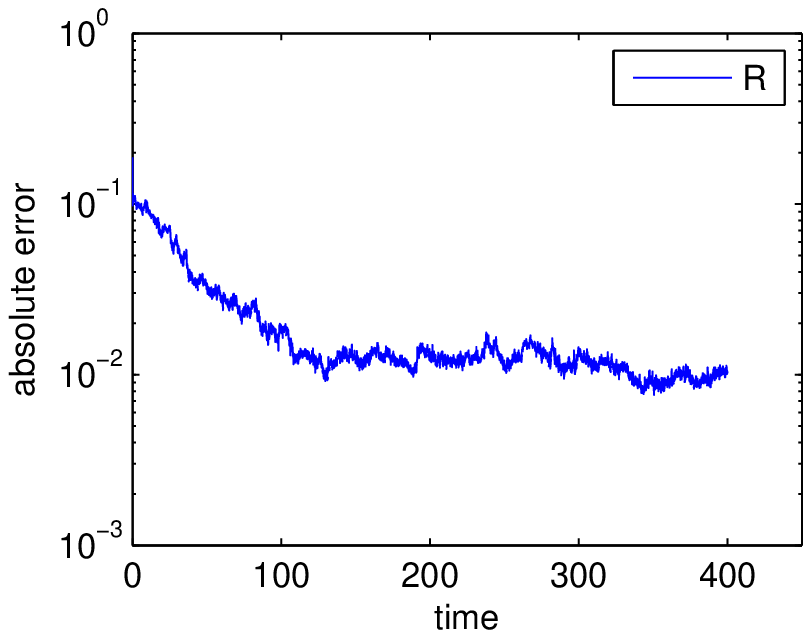}
        }
}
\centerline{
        \subfigure[Average absolute error of the posterior-mean estimates; parameter $B$.]{
                \includegraphics[width=0.35\linewidth]{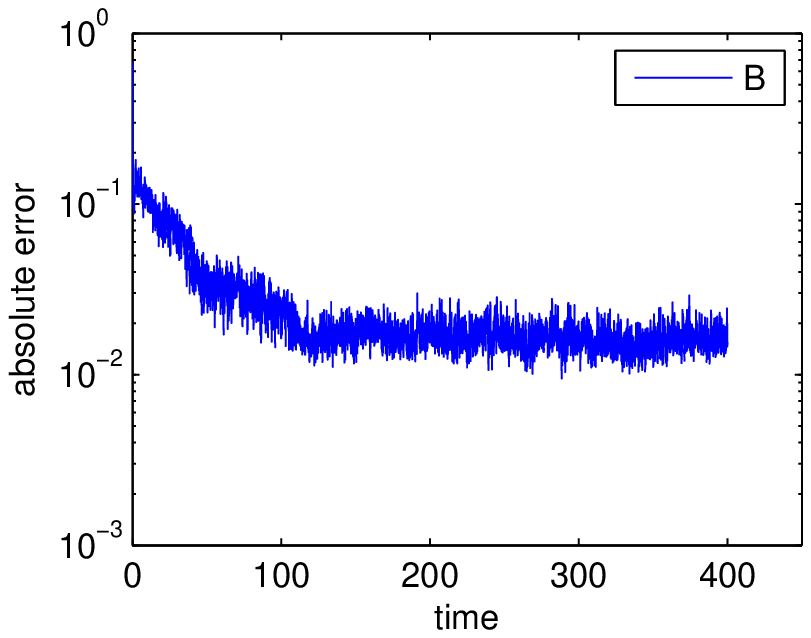}
        }
	\subfigure[Average absolute error of the posterior-mean estimates; parameter $k_o$.]{
                \includegraphics[width=0.35\linewidth]{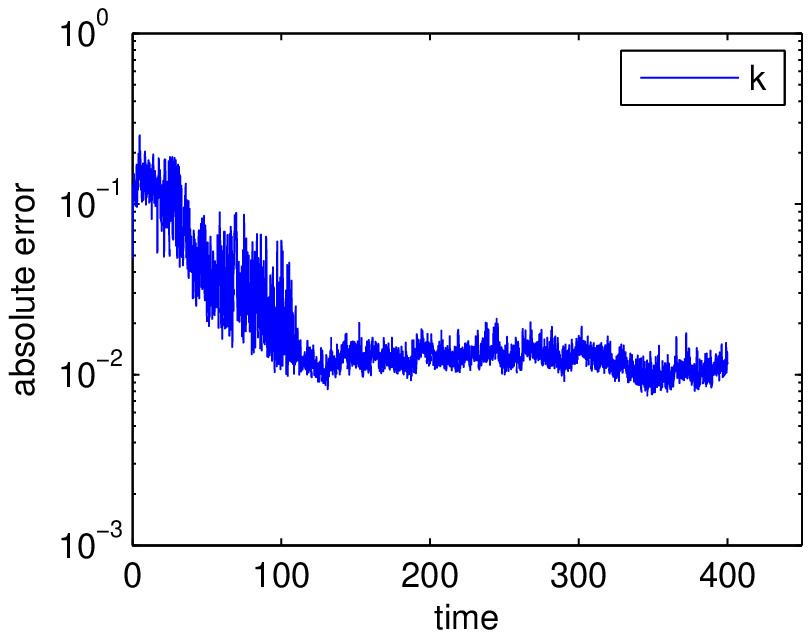}
        }
}
\caption{Absolute errors of the posterior-mean estimates of the Lorenz 63 model parameters, $S,R,B$ and $k_o$, versus continuous time. The errors have been averaged over 50 independent simulation runs. The length of each simulation is 400 continuous time units, which amounts to $400 \times 10^3$ discrete-time steps after discretisation of the Lorenz 63 model, with a sequence of 10,000 observations.}
\label{fMeanMSE}
\end{figure}

\begin{figure}
\centerline{
        \subfigure[Variance of the absolute error of the posterior-mean estimates; parameter $S$.]{
                \includegraphics[width=0.35\linewidth]{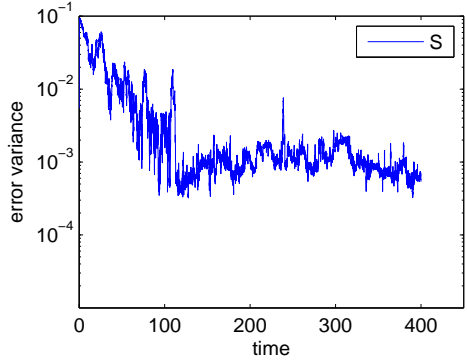}
        }
        \subfigure[Variance of the absolute error of the posterior-mean estimates; parameter $R$.]{
                \includegraphics[width=0.35\linewidth]{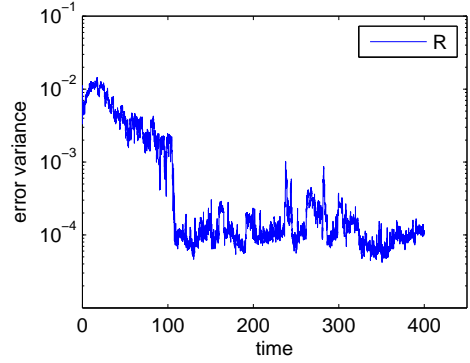}
        }
}
\centerline{
        \subfigure[Variance of the absolute error of the posterior-mean estimates; parameter $B$.]{
                \includegraphics[width=0.35\linewidth]{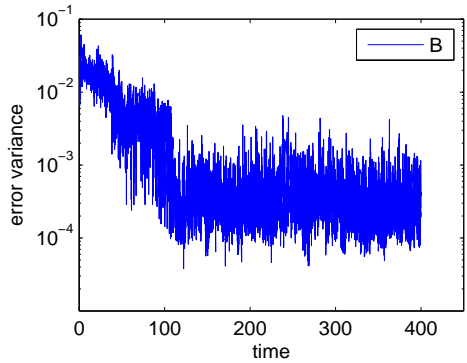}
        }
	\subfigure[Variance of the absolute error of the posterior-mean estimates; parameter $k_o$.]{
                \includegraphics[width=0.35\linewidth]{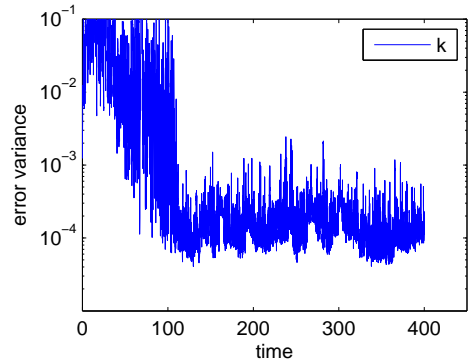}
        }
}
\caption{Variance of the absolute errors of the posterior-mean estimates of the Lorenz 63 model parameters, $S,R,B$ and $k_o$, versus continuous time. The variances have been estimated from 50 independent simulation runs. The length of each simulation is 400 continuous time units, which amounts to $400 \times 10^3$ discrete-time steps after discretisation of the Lorenz 63 model, with a sequence of 10,000 observations.}
\label{fVarMSE}
\end{figure}

Figure \ref{fMeanMSE} displays the mean absolute error for each parameter, $S,R,B$ and $k_o$, over time. We observe that there is a convergence period and, after approximately 100 continuous time units, the error converges to a steady value and remains stable for the rest of the simulation. The same kind of performance is observed for the variance of the absolute errors, computed over the same set of 50 independent simulations, and shown in Figure \ref{fVarMSE}.  

\section{Conclusions} \label{sConclusions}

We have analysed the asymptotic convergence of a recursive Monte Carlo scheme, consisting of two (nested) layers of particle filters, for the approximation and tracking of the posterior probability distribution of the unknown parameters of a state-space Markov system. The algorithm is similar to the recently proposed SMC$^2$ method, however the scheme in this paper is purely recursive and, thus, potentially more useful for online implementations.

The theoretical contribution of the paper includes the analysis of the errors in the approximation of integrals of bounded functions w.r.t. the posterior probability measure of the parameters. The analysis is carried out under regularity assumptions that include:
\begin{itemize}
\item The compactness of the parameter space.
\item The stability of the sequence of posterior probability measures of the unknown parameters, $\{ \mu_t \}$, w.r.t. the initial measure $\mu_0$.
\item A state space model that consists of a mixing Markov kernel and a normalised likelihood function with a positive lower bound. These regularity conditions are assumed to be satisfied uniformly over the parameter support. If this this assumption is met, then the classical results in \cite{DelMoral04} imply that the standard particle filters for the state space model of interest converge uniformly over time for any choice of the parameters in the support set $D_\theta$. 
\item The Markov kernel has a pdf (w.r.t. the Lebesgue measure) which is Lipschitz continuous w.r.t. the vector of unknown parameters. The likelihood function in the model is also assumed to be Lipschitz continuous w.r.t. the parameters.
\end{itemize}
These assumptions are restrictive, yet they simply describe a model for which the standard particle filter would converge uniformly over time (were the parameters known) {\em and} for which small perturbations to the parameters yield small perturbations in the sequence of posterior probability measures (for the same sequence of observations). The convergence of the proposed recursive algorithm cannot be guaranteed if any of the assumptions above is not met (e.g., for models in which some specific choice of the parameters may yield an unstable behaviour).

The uniform convergence result in Theorem \ref{thUniform} has additional implications. In this paper, we have proved that, for a class of non-ambiguous models \cite{Papavasiliou06}, the parameters can be identified, i.e., they can be estimated in an asymptotically exact manner (meaning that the sequence of approximate posterior measures generated by the algorithm converge to a delta measure). 
 
\section*{Acknowledgements} 

The work of J. M\'{\i}guez was partially supported by {\em Ministerio de Econom\'{\i}a y Competitividad} of Spain (project TEC2012-38883-C02-01 COMPREHENSION) and the Office of Naval Research Global (award no. N62909-15-1-2011). Part of this work was carried out while J. M. was a visitor at the Department of Mathematics of Imperial College London, with partial support from an EPSRC Mathematics Platform grant. D. C. and J. M. would also like to acknowledge the support of the Isaac Newton Institute through the program ``Monte Carlo Inference for High-Dimensional Statistical Models''.


\appendix

\section{A proof for inequality \eqref{eqPasito1}} \label{apMus}

We need to prove that $\| (v,\bar \mu_{n-1}^N) - (v,\mu_{n-1}^N) \|_p \le \frac{s_1\| v \|_\infty}{\sqrt{N}}$ for some $s_1<\infty$ independent of $N$ and $v \in B(D_\theta)$. 

Recall that we draw the particles $\bar \theta_n^{(i)}$, $i=1, \ldots, N$, independently from the  kernels $\kappa_{N,\sfp}^{\theta_{n-1}^{(i)}}$, $i=1, \ldots, N$, respectively, and start from the triangle inequality 
\begin{equation}
\| (v,\bar \mu_{n-1}^N) - (v,\mu_{n-1}^N) \|_p \le 
\| (v,\bar \mu_{n-1}^N) - (v, \kappa_{N,\sfp}\mu_{n-1}^N) \|_p + \| (v,\kappa_{N,\sfp}\mu_{n-1}^N) - (v, \mu_{n-1}^N) \|_p 
\label{eqRetriangle}
\end{equation}
where
$$
(v,\kappa_{N,\sfp}\mu_{n-1}^N) = \frac{1}{N} \sum_{i=1}^N \int v(\theta) \kappa_{N,\sfp}^{\theta_{n-1}^{(i)}}(d\theta),
$$
and then analyse the two terms on the right hand side of \eqref{eqRetriangle} separately. 

Let $\mG_{n-1}$ be the $\sigma$-algebra generated by the random particles 
$\{ \bar \theta_{1:n-1}^{(i)}, \theta_{0:n-1}^{(i)} \}_{1 \le i \le N}$. 
Then 
\begin{equation}
E\left[
	(v,\bar \mu_{n-1}^N) | \mG_{n-1}
\right] = 
\frac{1}{N} \sum_{i=1}^N \int v(\theta)\kappa_{N,\sfp}^{\theta_{n-1}^{(i)}}(d\theta) = (v,\kappa_{N,\sfp}\mu_{n-1}^N)
\nonumber
\end{equation}
and the difference $(v,\bar \mu_{n-1}^N) - (v,\kappa_{N,\sfp}\mu_{n-1}^N)$ can be written as
\begin{equation}
(v,\bar \mu_{n-1}^N) - (v,\kappa_{N,\sfp}\mu_{n-1}^N) = \frac{1}{N} \sum_{i=1}^N \bar Z_{n-1}^{(i)},
\nonumber
\end{equation}
where the random variables $\bar Z_{n-1}^{(i)} = v(\bar \theta_n^{(i)}) - E[ v(\bar \theta_n^{(i)}) | \mG_{n-1} ]$, $i=1, ..., N$, are conditionally independent (given $\mG_{n-1}$), have zero mean and can be bounded as $| \bar Z_{n-1}^{(i)} | \le 2\| v \|_\infty$. As a consequence, it is an exercise in combinatorics to show that
\begin{equation}
E\left[
	\left|
		(v,\bar \mu_{n-1}^N) - (v,\kappa_{N,\sfp}\mu_{n-1}^N)
	\right|^p | \mG_{n-1}
\right] = E\left[
	\left|
		\frac{1}{N} \sum_{i=1}^N \bar Z_{n-1}^{(i)}
	\right|^p | \mG_{n-1}
\right] \le \frac{
	\tilde c_{1}^p \| v \|_{\infty}^p
}{
	N^\frac{p}{2}
}, \label{eqBoundingFirstTerm-1}
\end{equation}
where $\tilde c_{1}$ is a constant independent of $N$, $n$ and $v$ (actually, independent of the distribution of the $\bar Z_{n-1}^{(i)}$'s).  From \eqref{eqBoundingFirstTerm-1} we readily obtain that
\begin{equation}
\| (v,\bar \mu_{n-1}^N) - (v,\kappa_{N,\sfp}\mu_{n-1}^N) \|_p \le \frac{
	\tilde c_{1}\| v \|_\infty
}{
	\sqrt{N}
}.
\label{eqBoundingFirstTerm-2}
\end{equation}

For the remaining term in \eqref{eqRetriangle}, namely, $\| (v,\kappa_{N,\sfp}\mu_{n-1}^N) - (v,\mu_{n-1}^N)\|_p$, we simply note that
\begin{eqnarray}
\left| 
	(v,\kappa_{N,\sfp}\mu_{n-1}^N) - (v,\mu_{n-1}^N)
\right| &=& \left|
	\frac{1}{N} \sum_{i=1}^N \int \left(
		v(\theta) - v(\theta_{n-1}^{(i)})
	\right) \kappa_{N,\sfp}^{\theta_{n-1}^{(i)}}(d\theta)
\right| \nonumber \\
&\le& \frac{1}{N} \sum_{i=1}^N \int \left|
	v(\theta) - v(\theta_{n-1}^{(i)})
\right| \kappa_{N,\sfp}^{\theta_{n-1}^{(i)}}(d\theta) 
\le \frac{
	2 \|v\|_\infty
}{
	\sqrt{N}
},
\label{eqBoundingSecond-1}
\end{eqnarray}
where the last inequality follows from Proposition \ref{propAssumptionKappa1}. 

Substituting the inequalities \eqref{eqBoundingFirstTerm-2} and \eqref{eqBoundingSecond-1} into Eq. \eqref{eqRetriangle} yields the desired conclusion, viz., Eq. \eqref{eqPasito1}, with constant $s_1 = 2 + \tilde c_1$ independent of $N$.

\bibliographystyle{plain}
\bibliography{bibliografia}
\end{document}